\documentclass[a4paper,11pt]{article}
\usepackage{tocloft}
\pdfoutput=1
\usepackage{jheppub}
\usepackage{amssymb,amsmath,mathrsfs,enumerate,dsfont}
\usepackage{graphicx,rotate,multicol}
\usepackage{float}
\usepackage{subcaption}
\usepackage[export]{adjustbox}
\usepackage{braket}

\usepackage{slashed}
\usepackage{mathtools}
\usepackage{multirow}
\usepackage{bbold}
\usepackage{orcidlink}

\allowdisplaybreaks

\title{\boldmath Stodolsky effect in the framework of Generalised Neutrino Interactions}

\author[a,b]{Siddhartha Bandyopadhyay,$^{\orcidlink{0009-0002-3455-643X},}$}
\author[b]{Ujjal Kumar Dey$^{\orcidlink{0000-0002-9620-7561},}$}
\affiliation[a]{Indian Institute of Technology Kanpur,\\Kalyanpur, Kanpur 208016, Uttar Pradesh, India}
\affiliation[b]{Department of Physical Sciences, Indian Institute of Science Education and Research Berhampur,\\Ganjam, Odisha, 760003, India}

\emailAdd{siddhartha25@iitk.ac.in}
\emailAdd{ujjal@iiserbpr.ac.in}

\abstract{We study the Stodolsky effect utilizing the most general form of neutrino interactions with electrons below the electroweak scale by considering all possible Lorentz invariant operators respecting SU(3)$\otimes$U(1) symmetry. We perform our calculation for both Dirac and Majorana neutrinos and find that in the most general setting, only the non-standard neutrino interactions and the tensor interaction terms provide a non-zero contribution, apart from the Standard Model contribution. We investigate the implications for the possible detection of the cosmic neutrino background (C$\nu$B) by analysing the energy shifts that are characteristic of the Stodolsky effect. We also discuss the implication of considerable asymmetry in the C$\nu$B on the present scenario.}

\begin{document}
\maketitle
\flushbottom


\section{Introduction}
\label{sec:intro}
Neutrinos are one of the most prominent reminders of the incompleteness of the otherwise highly successful Standard Model (SM) of particle physics. It also opens the deliberations on the physics Beyond Standard Model (BSM). Experimentally observed oscillation phenomena of neutrinos necessitate that they possess non-vanishing masses which is in dissonance with the SM. To address this, augmenting SM with additional fields and/or gauge symmetries has been the standard BSM lore for decades. Instead of chasing each and every such scenario, which most of the time are untenable due to the exorbitant characteristic energy scales, it is much more practical and economical to parametrize possible BSM interactions in terms of a low-energy
effective field theory.    
One can formulate the most general interactions and try to constrain the parameters using data obtained from experiments. Non-standard neutrino interactions (NSIs) are a well-known framework for parametrizing BSM effects in the neutrino sector~\cite{Miranda_2015}. Recently, more exotic interactions like scalar, pseudoscalar, vector, axial-vector, and tensor structures have been taken into account. Together with the NSIs, they constitute the generalised neutrino interactions (GNIs)~\cite{BISCHER19, Bergmann99, Flore:2026qeh}. 
Previous experimental collaborations and simulations for future efforts such as DUNE \cite{BISCHERDUNE} provides bounds on the NSI \cite{NSI_status, Khan_2017, freitas2025, Davidson_2003, coloma2023, CPNova, NovaT2K, Gehrlein_2025, BOREXINO:2026owb} and GNI parameters \cite{Amir_Rode, Escrihuela:2021mud, Banerjee:2023lrk}. The constraining of tensor parameters from the neutrino magnetic moment measurements is an avenue that has been explored as well \cite{Healey:2013vka,Sevda:2016otj}. 
In the context of modern cosmology, various properties of neutrinos are of immense importance. The detection of the cosmic neutrino background (C$\nu$B), predicted by standard cosmological theories, is one of the most challenging aspects of modern cosmology and is the holy grail of neutrino physics. A number of direct and indirect detection proposals have been made over the years \cite{Opher:1974drq, Lewis1980, Cabibbo:1982bb, Shvartsman:1982sn,  Ringwald:2004np, Eberle2004, Graciela2005, PauliBlocking,  Shergold_2021, Bauer2021, Bauer_2023, delCastillo:2025qnr, Herrera:2026pzj}. The PTOLEMY experiment \cite{PTOLEMY1, PTOLEMY2, Virzi:2024} aims at directly detecting relic neutrinos by the observation of inverse beta decays, a method first proposed by S. Weinberg \cite{Weinberg62}. However, further research has led to the indication that PTOLEMY may not be viable for the detection of relic neutrinos, being plagued by the uncertainty principle \cite{Boyarsky, Zohar}. The possible ways in which the quantum uncertainty can be addressed is discussed in \cite{PTOCollab}. In recent times, attention has shifted towards the elastic scattering of relic neutrinos on macroscopic targets. This can be roughly divided into coherent neutrino scattering and the Stodolsky effect. Coherent Elastic Neutrino-Nucleus Scattering, popularly known as CE$\nu$NS, is a $\mathcal{O}(G^2_F)$ effect that deals with the recoil acceleration of atoms in a material due to the neutral current interaction of the relic neutrinos with the nucleons and electrons present in the atoms \cite{RevModPhys.84.1307, Domcke_2017}. The effect picks up an additional enhancement factor as the coherence length of the relic neutrinos are spread over the expanse of many nuclei in the solid lattice \cite{Shergold_2021}. The scattered neutrinos impart an average momentum transfer which induces a microscopic acceleration of the target material. The Stodolsky effect is a $\mathcal{O}(G_F)$ effect \cite{Stodolsky:1974aq, Duda_2001, Domcke_2017, darkstod} which arises due to the interaction of relic neutrinos with atomic electrons causing a change in the energy levels of different spin states. This energy splitting between electrons of spin $+\frac{1}{2}$ and $-\frac{1}{2}$ leads to a small torque on the electron which is proportional to the energy shift -- an effect very similar to that of the Zeeman effect. As a result, a ferromagnetic material must experience a microscopic torque which in principle could be measured using a torsion balance \cite{Bauer_2023,Li:2026rty,Kalia:2024eml}. Ferromagnetic gyroscopes \cite{Fadeev:2020xzw} could serve as promising detectors as well.
In this work we focus our attention to the Stodolsky effect and derive the expression for the energy shift of atomic electronic states considering the most general form of neutrino interactions up to dimension six operators. This serves as a generalisation of existing formulae in the literature \cite{Stodolsky:1974aq,Duda_2001,Graciela2005,Domcke_2017,Bauer_2023} and allows us to estimate the energy splitting using constraints on the parameter space of BSM interactions.
The paper is organised as follows. In Sec. \ref{sec:GNI} we provide a brief outlook on GNIs. For our study of the Stodolsky effect, we are concerned only with the interactions of neutrinos involving electrons. In Sec. \ref{sec:Stodolsky} we start with a brief introduction to the Stodolsky effect followed by the derivation of the energy shift using the interaction Lagrangian. We then discuss our results in the context of C$\nu$B, for both the standard cosmological densities and  in the case of overdensities. We also extend our formalism to flavor eigenstate neutrinos. We finally summarise and conclude in Sec. \ref{sec:sumandconc}.

\section{Generalised Neutrino Interactions}
\label{sec:GNI}

To set the notations and conventions, in this section we briefly review the generalised neutrino interactions (GNI) which extends the idea of non-standard neutrino interaction (NSI) by augmenting it with additional interaction possibilities. Since we are interested in the description of Stodolsky effect the effective interactions involving electrons are going to be crucial. The most general dimension-6 effective operators maintaining $\text{SU(3)}_\text{c} \otimes \text{U(1)}_\text{em}$ symmetry relevant for the neutrino-electron interaction can be written as~\cite{BISCHERDUNE},
\begin{equation}
\label{GNILag}
	\mathcal{L}=-\frac{G_\text{F}}{\sqrt{2}}\sum_{\alpha,\beta} \sum^{10}_{k=1}\: {\stackrel{(\sim)k}{\epsilon_{\alpha \beta}}}\:(\bar{\nu}_\alpha \mathcal{O}_k \nu_\beta)\,(\bar{e}\mathcal{O}^\prime_k e)\,,
\end{equation}
where, $\epsilon^k$ and $\tilde{\epsilon}^k$ are the dimensionless coupling constants of the GNI Lagrangian. The different operators and the corresponding couplings are listed in Table \ref{table:Tab1}. In the Standard Model (SM), all the parameters except ${\epsilon}^L$ and ${\epsilon}^R$ (see Table  \ref{table:Tab1}) are zero such that,
\begin{equation}
	\epsilon^{L,SM}_{\alpha\beta}=\delta_{\alpha e}\delta_{\beta e}+\biggl(-\frac{1}{2}+\sin^2 \theta_W \biggl)\delta_{\alpha \beta} ~,~~~\epsilon^{R,SM}_{\alpha\beta}=\sin^2 \theta_W \delta_{\alpha \beta}\,,
\end{equation}
where $\theta_W$ is the Weinberg angle. As such we define $\epsilon^{(L/R)}=\epsilon^{{(L/R)},{\text{SM}}}+\epsilon^{(L/R),\text{NSI}}$. In general, the Lagrangian Eq. (\ref{GNILag}) must contain the Hermitian conjugate operators as well. Mathematically, adding the Hermitian conjugate part is equivalent to imposing the following restrictions on the GNI parameters,
\begin{align}
\label{Constraints}
	&\epsilon^L_{\alpha\beta}=\epsilon^{L*}_{\beta\alpha},~~~\tilde{\epsilon}^L_{\alpha\beta}=\tilde{\epsilon}^{L*}_{\beta\alpha},~~~\epsilon^R_{\alpha\beta}=\epsilon^{R*}_{\beta\alpha},~~~\tilde{\epsilon}^R_{\alpha\beta}=\tilde{\epsilon}^{R*}_{\beta\alpha}\,,\notag \\
	&\epsilon^S_{\alpha\beta}=\tilde{\epsilon}^{S*}_{\beta\alpha},~~~\epsilon^P_{\alpha\beta}=-\tilde{\epsilon}^{P*}_{\beta\alpha},~~~\epsilon^T_{\alpha\beta}=\tilde{\epsilon}^{T*}_{\beta\alpha}~.
\end{align}
In the most general setting, the parameters are unconstrained and the neutrinos are modeled to be of the Dirac type. The number of free parameters are greatly reduced if we impose CP symmetry (in which case the parameters are all real) and/or consider neutrinos to be of the Majorana type \cite{Rosen1982, Rodejohann_2017, BISCHERDUNE}. In the latter case we have,
\begin{align}
\label{MajoranaConstraints}
	\epsilon^L_{\alpha\beta}=-\tilde{\epsilon}^L_{\beta\alpha},~~~\epsilon^R_{\alpha\beta}=-\tilde{\epsilon}^R_{\beta\alpha},~~~\epsilon^S_{\alpha\beta}=\epsilon^S_{\beta\alpha},~~~\epsilon^P_{\alpha\beta}=\epsilon^P_{\beta\alpha},~~~\epsilon^T_{\alpha\beta}=-\epsilon^T_{\beta\alpha}~.
\end{align}
By allowing the Lorentz invariant structures listed in Table \ref{table:Tab1}, we incorporate the possible existence of right handed neutrinos in our analysis. Although there is no concrete evidence for them yet, one can still constrain their effective interactions with SM electrons in a model independent way. Previously, bounds had been set on the scalar, pseudoscalar and tensor parameters from the CHARM-II collaboration \cite{VILAIN1993351, VILAIN1994246} and NSI bounds from the Borexino detector \cite{Khan_2017}. Recently, bounds on NSI parameters using actual NO$\nu$A data have been derived \cite{Gehrlein_2025}. The expected bounds for GNI parameters have been estimated for DUNE \cite{BISCHERDUNE}, Borexino \cite{Amir_Rode, BOREXINO:2026owb} and recently FASER \cite{Escrihuela:2023sfb}.

\begin{table}

\renewcommand{\arraystretch}{1.2}
\centering
\begin{tabular}{l c c l}
\hline
$k$~~ & $\tilde{\epsilon}^k$ & $\mathcal{O}_k$ &~~ $\mathcal{O}^\prime_k$  \\
\hline
1& $\epsilon^L$ & $\gamma_\mu(\mathds{1}-\gamma^5)$ & $\gamma_\mu(\mathds{1}-\gamma^5)$\\
2& $\tilde{\epsilon}^L$ & $\gamma_\mu(\mathds{1}+\gamma^5)$ & $\gamma_\mu(\mathds{1}-\gamma^5)$\\
3& $\epsilon^R$ & $\gamma_\mu(\mathds{1}-\gamma^5)$ & $\gamma_\mu(\mathds{1}+\gamma^5)$\\
4& $\tilde{\epsilon}^R$ & $\gamma_\mu(\mathds{1}+\gamma^5)$ & $\gamma_\mu(\mathds{1}+\gamma^5)$\\
5& $\epsilon^S$ & $(\mathds{1}-\gamma^5)$ & ~~~~~$\mathds{1}$\\
6& $\tilde{\epsilon}^S$ & $(\mathds{1}+\gamma^5)$ & ~~~~~$\mathds{1}$\\
7& $-\epsilon^P$ & $(\mathds{1}-\gamma^5)$ & ~~~~~$\gamma^5$\\
8& $-\tilde{\epsilon}^P$ & $(\mathds{1}+\gamma^5)$ & ~~~~~$\gamma^5$\\
9& $\epsilon^T$ & $\sigma_{\mu\nu}(\mathds{1}-\gamma^5)$ & $\sigma^{\mu\nu}(\mathds{1}-\gamma^5)$\\
10& $\tilde{\epsilon}^T$ & $\sigma_{\mu\nu}(\mathds{1}+\gamma^5)$ & $\sigma^{\mu\nu}(\mathds{1}+\gamma^5)$\\

\hline

\end{tabular}
\caption{The different coupling constants and their corresponding operators appearing in the GNI Lagrangian Eq. (\ref{GNILag}). }
\label{table:Tab1}
\end{table}

\section{Stodolsky effect}
\label{sec:Stodolsky}
In this section we briefly describe the basics of the Stodolsky effect. In the presence of a steady neutrino background, electrons in a given volume of space experience small energy shifts which depend on their spin eigenvalues. Stodolsky \cite{Stodolsky:1974aq} pointed out that in the presence of a cosmic neutrino background two kinds of phenomena can take place -- (i) the direction of the spin vector of a transversely polarized moving electron will rotate, and (ii) the motion of the earth through the rest frame of the C$\nu$B causes the development of a torque on a ferromagnetic test object. Both of these effects depend on the energy shift of the electron which, in the electron rest frame depends on its velocity through the C$\nu$B rest frame (the frame in which neutrino velocities are isotropic) and neutrino number density asymmetries \cite{hagmann1999cosmic}. Traditional calculations of the energy shift use the tree-level neutrino-electron four-fermion interaction term which can be derived as a low energy effective field theoretic description of the SM obtained after integrating out the $Z$ and $W$ bosonic degrees of freedom. The interaction part of the Lagrangian then becomes,
\begin{equation}
\label{Shergold1}
\mathcal{L}=-\mathcal{H}=-\frac{G_F}{\sqrt{2}}\Bigl[\Bar{\nu}_e \gamma^\mu(1-\gamma^5)\nu_e\,\Bar{e}\gamma_\mu(1-\gamma^5)e+\sum_\alpha \Bar{\nu}_\alpha \gamma^\mu(1-\gamma^5)\nu_\alpha \,\Bar{e}\gamma_\mu(g^e_V-g^e_A \gamma^5)e   \Bigl],
\end{equation}
where $\alpha$ runs over the neutrino flavor indices.
Using the notation of traditional $V-A$ theory we denote $g^e_V=-1/2+2\sin^2 \theta_W$ and $g^e_A=-1/2$, with $\theta_W$ denoting the Weinberg mixing angle \cite{Sudarshan1958, Feynman1958}. Since we are developing our formalism in the context of the C$\nu$B it is customary to work in the neutrino mass eigenbasis. We will discuss relic neutrinos in detail in Sec. \ref{subsec:CnuB}. The leading order contribution to the energy shift of an electron with given 3-momentum $\vec{p}_e$ and spin $s_e$ due to the interaction with neutrinos/anti-neutrinos ($\nu_i/\Bar{\nu}_i$, $i=1,2,3$) of approximately uniform 3-momentum $\vec{p}_{\nu_i}$ (considering it to be the same for neutrinos and anti-neutrinos) and corresponding helicity $s$ can be written as,
\begin{equation}
	\label{Energyshift1}
	\Delta E_e(\vec{p}_e,s_e)=\sum_{\nu,i,s}\sum_{N_\nu} \langle e_{p_e,s_e},\nu_{p_{\nu_i},s}|:\int d^3x \,\mathcal{H}(x):|e_{p_e,s_e},\nu_{p_{\nu_i},s}\rangle\,,
\end{equation}
where the sum is taken over all background neutrinos $N_\nu$ of a kind ($\nu,i,s$) in a given region of space $V$, neutrinos and anti-neutrinos,
three mass eigenstates and two helicities.
This energy shift is defined in the lab frame and subsequently, all the dynamical quantities are measured in this frame as well. The detailed calculation of the energy shifts is given in Appendix \ref{appendixA}.
Since, our aim is to find the energy difference between the two electronic spin states, we can simplify our expressions noting that terms in the energy shift that do not depend upon the spin of the electron cancel out anyway. Then we can write down the energy shift for an electron at rest in the lab frame due to the interaction with mass eigenstate Dirac neutrinos as,
\begin{align}\label{penE}
	&\Delta E^D_e (\vec{0},s_e)=\frac{G_F}{\sqrt{2}}\sum_{i,s}\Biggr[\Bigl(n_{\nu}(\nu^D_{i,s})-n_\nu ({\Bar{\nu}}^D_{i,s})\Bigl)\Biggl\{(-\epsilon^L_{ii}-\tilde{\epsilon}^L_{ii}+\epsilon^R_{ii}+\tilde{\epsilon}^R_{ii})\biggl\langle\frac{S_e \cdot p_{\nu_i}}{E_{p_{\nu_i}}} \biggl\rangle\notag\\ &+\frac{4}{m_e}\biggl(\biggl\langle\frac{(S_{\nu_i}\cdot p_e)(S_e \cdot p_{\nu_i})}{E_{p_{\nu_i}}} \biggl\rangle(\epsilon^T_{ii}+\tilde{\epsilon}^T_{ii})-\biggl\langle\frac{(p_e \cdot p_{\nu_i})(S_e \cdot S_{\nu_i})}{E_{p_{\nu_i}}} \biggl\rangle (\epsilon^T_{ii}+\tilde{\epsilon}^T_{ii})\notag\\
&+i\biggl\langle\frac{\epsilon_{\alpha\beta\mu\nu}p^\alpha_e S^\beta_e p^\mu_{\nu_i} S^\nu_{\nu_i}}{E_{p_{\nu_i}}} \biggl\rangle (\epsilon^T_{ii}-\tilde{\epsilon}^T_{ii})+ \frac{1}{4}\biggl\langle \frac{(p_{\nu_i}\cdot S_{\nu_i}) (p_e \cdot S_e)} {E_{p_{\nu_i}} } \biggl\rangle(\epsilon^P_{ii}-\tilde{\epsilon}^P_{ii})\biggl)\Biggl\}\notag\\
&+\Bigl(n_{\nu}(\nu^D_{i,s})+n_\nu ({\Bar{\nu}}^D_{i,s})\Bigl)\Biggl\{(\epsilon^L_{ii}-\tilde{\epsilon}^L_{ii}-\epsilon^R_{ii}+\tilde{\epsilon}^R_{ii})m_{\nu_i}
\biggl\langle\frac{S_e \cdot S_{\nu_i}}{E_{p_{\nu_i}}} \biggl\rangle+(\epsilon^P_{ii}+\tilde{\epsilon}^P_{ii})\frac{m_{\nu_i}}{m_e}\biggl\langle\frac{p_e \cdot S_e}{E_{p_{\nu_i}}} \biggl\rangle \Biggl\}\Biggr]\notag\\ &+\text{terms independent of}\:{S_e}\,,
\end{align}
\newline
and for Majorana neutrinos as,
\begin{align}\label{penEM}
	\Delta E^M_e (\vec{0},s_e)&={\sqrt{2}}{G_F}\sum_{i,s}\Biggr[n_{\nu}(\nu^D_{i,s})\Biggl\{(\epsilon^L_{ii}-\tilde{\epsilon}^L_{ii}-\epsilon^R_{ii}+\tilde{\epsilon}^R_{ii})m_{\nu_i}\biggl\langle\frac{S_e \cdot S_{\nu_i}}{E_{p_{\nu_i}}} \biggl\rangle\notag\\ &+\frac{m_{\nu_i}}{m_e}(\epsilon^P_{ii}+\tilde{\epsilon}^P_{ii})\biggl\langle\frac{p_e\cdot S_e}{E_{p_{\nu_i}}} \biggl\rangle +\frac{1}{m_e}(\epsilon^P_{ii}-\tilde{\epsilon}^P_{ii})\biggl\langle\frac{(p_{\nu_i}\cdot S_{\nu_i})(p_e\cdot S_e)}{E_{p_{\nu_i}}} \biggl\rangle \Biggl\}\Biggr]\notag \\
	&+\text{terms independent of}\:{S_e}\,,
\end{align}
where $\langle \; \rangle$ denotes flux averaging (see Eq. (\ref{Fluxavg})). Note that the scalar couplings $\epsilon^S$ and $\tilde{\epsilon}^S$ are present in the terms independent of the electronic spin and thus do not appear when we take the energy splitting between the two electronic spin states. Moreover, from the definition of the spin 4-vector (see Appendix \ref{appendixA}) we can easily derive that $p\cdot S=p^\mu S_\mu=0$ and $\epsilon_{\alpha\beta\gamma\delta}p^\alpha_e S^\beta_e p^\gamma_{\nu_i} S^\delta_{\nu_i}=0$. This causes one tensor coupling term to vanish and all pseudoscalar couplings to completely vanish, so that we are left with,
\begin{align}\label{finalE}
	\Delta E^D_e (\vec{0},s_e)&=\frac{G_F}{\sqrt{2}}\sum_{i,s}\Biggr[\Bigl(n_{\nu}(\nu^D_{i,s})-n_\nu ({\Bar{\nu}}^D_{i,s})\Bigl)\Biggl\{(-\epsilon^L_{ii}-\tilde{\epsilon}^L_{ii}+\epsilon^R_{ii}+\tilde{\epsilon}^R_{ii})\biggl\langle\frac{S_e \cdot p_{\nu_i}}{E_{p_{\nu_i}}} \biggl\rangle\notag\\ &+\frac{4}{m_e}(\epsilon^T_{ii}+\tilde{\epsilon}^T_{ii})\biggl(\biggl\langle\frac{(S_{\nu_i}\cdot p_e)(S_e \cdot p_{\nu_i})}{E_{p_{\nu_i}}} \biggl\rangle-\biggl\langle\frac{(p_e \cdot p_{\nu_i})(S_e \cdot S_{\nu_i})}{E_{p_{\nu_i}}} \biggl\rangle\biggl) \Biggl\}\notag\\
&+\Bigl(n_{\nu}(\nu^D_{i,s})+n_\nu ({\Bar{\nu}}^D_{i,s})\Bigl)\Biggl\{(\epsilon^L_{ii}-\tilde{\epsilon}^L_{ii}-\epsilon^R_{ii}+\tilde{\epsilon}^R_{ii})m_{\nu_i}
\biggl\langle\frac{S_e \cdot S_{\nu_i}}{E_{p_{\nu_i}}} \biggl\rangle \Biggl\}\Biggr]\notag \\
&+\text{terms independent of}\:{S_e}\,.
\end{align}
Thus, the energy shift for Dirac neutrinos depends only on the SM, NSI and tensor parameters. It also depends upon the number density of neutrinos and anti-neutrinos of a given helicity. For Majorana neutrinos we only have contribution from the SM and NSI parameters such that,
\begin{align}\label{finalEM}
	\Delta E^M_e (\vec{0},s_e)&={\sqrt{2}}{G_F}\sum_{i,s}\Biggr[n_{\nu}(\nu^D_{i,s})\Biggl\{(\epsilon^L_{ii}-\tilde{\epsilon}^L_{ii}-\epsilon^R_{ii}+\tilde{\epsilon}^R_{ii})m_{\nu_i}\biggl\langle\frac{S_e \cdot S_{\nu_i}}{E_{p_{\nu_i}}} \biggl\rangle \Biggl\}\Biggr]\notag \\
	&+\text{terms independent of}\:{S_e}\,.
\end{align}
We note that Eq.~(\ref{finalE}) can be reorganised so that the energy shift can be written as $\Delta E = {\Delta E}_{\text{SM}}+{\Delta E}_{\text{NSI}}+{\Delta E}_{\text{Tensor}}$. We find that ${\Delta E}_{\text{SM/NSI}} \sim \frac{G_F}{\sqrt{2}}\Delta n$ and ${\Delta E}_{\text{Tensor}}\sim \frac{G_F}{\sqrt{2}}\Delta n+\sqrt{2}{G_F}n_{\nu}$ upto appropriate flux averages and couplings, where $\Delta n=\eta_\nu \times n_{\gamma}$ is the difference of number densities between neutrinos and anti-neutrinos. Hence, we find that the energy shift due to the tensor couplings depends not only on the asymmetry between neutrino number densities but also on the number densities themselves. This phenomenon is absent in the SM and NSI case. For $\eta_{\nu}=0$ the energy shift contribution from the SM and NSI sectors vanish leaving behind the only non-zero contribution from the tensor sector.

Since the energy shift of the electron is dependent on the spin of the electron itself, the spin operators along directions orthogonal to the propagation of neutrinos and the perturbed Hamiltonian of the electron no longer commute which leads to a net torque $\tau_e \simeq \Delta E_e$. Thus, if one takes a ferromagnet with $N_e$ polarised electrons, the total torque due to the neutrino flux incident upon the material is \cite{Bauer_2023},
\begin{equation}\label{ferrotorque}
	N_e \tau_e \simeq \frac{N_A}{m_A}\frac{n_e}{A}M|\Delta E_e|\,,
\end{equation}
where $N_A$ is Avogadro's constant, $n_e$ is the average number of unpaired electrons per atom, $A$ is the mass number of the target material, respectively, $M$ is the total mass of the target material, and $m_{A}=1 \text{g mol}^{-1}$. If we consider a ferromagnet, with spatial dimension $R$ and moment of inertia $I\simeq M R^2$, kept in front of the incoming flux of neutrino particles, there will be a linear acceleration given by,
\begin{equation}
	\label{ferroacc}
	a\simeq \frac{N_A}{m_A}\frac{n_e}{A R}|\Delta E_e|\,.
\end{equation}
In a similar but different kind of effect, one can consider a beam of transversely polarized electrons which is a coherent linear combination of the two possible spin states ($\pm$1) \cite{Swartz:1987xme}. While moving through the neutrino medium, the two spin states are split due to the interaction affecting the states differently which leads to the development of an overall phase between the two components of the transversely polarized state. This leads to a rotation of the polarization along the line of flight \cite{Stodolsky:1974aq}. This phase difference between the two components is given in the electron rest frame by,
\begin{equation}
	\label{optrot}
	\Delta\varphi\simeq\Delta E_e \Delta t\,,
\end{equation}
where $\Delta E_e$ is the energy splitting and $\Delta t$ is the time elapsed, both in the rest frame of the electron. Recently, the possible detection of such a phase difference has been discussed in the context of matter interferometers using a beam of fermionic atoms \cite{nugroho2025}. In this paper we will not look at this phenomenon and only restrict ourselves to lab frame targets bombarded with relic neutrinos. One can, in principle, try to measure the physical quantities given by Eqs. (\ref{ferroacc}) and (\ref{optrot}) and try to constrain the energy splitting. Conversely, new physics effects arising from an EFT Lagrangian of the form Eq. (\ref{GNILag}) can be probed by looking at the variation of these physical quantities from the SM values. 
The flux averages in Eq. (\ref{finalE}) depend upon the momentum distribution functions of the neutrinos and thus are different for neutrinos of different origin. In the following subsections, we take a look at relic neutrinos and estimate the value of the energy splitting and formulate the Stodolsky effect for flavor neutrinos.

\subsection{Cosmic Neutrino Background}
\label{subsec:CnuB}
In this section, we take a look at the relic neutrinos, which are predicted to constitute the C$\nu$B and estimate the energy splitting due to such neutrinos. In order to have a correct estimation of the energy splitting between the two electronic states ($\Delta E_e(s_e=+1)-\Delta E_e(s_e=-1)$) one needs to have the neutrino and anti-neutrino number densities along with the knowledge of the kinematics of the neutrino flux involved. Only then can one perform the flux averages in Eqs. (\ref{finalE}) and (\ref{finalEM}). 
For the study of relic neutrinos, one generally defines a C$\nu$B frame in which the relic neutrino velocities are isotropic i.e. the momentum vector of any such neutrino is given by,
\begin{equation}
	\vec{\tilde{p}}_{\nu_i}=|\vec{\tilde{p}}_{\nu_i}|(\text{cos}\,\tilde{\phi}\,\text{sin}\,\tilde{\theta}\,\hat{x}+\text{sin}\,\tilde{\phi}\,\text{sin}\,\tilde{\theta}\,\hat{y}+\text{cos}\,\tilde{\theta}\,\hat{z})\,,
\end{equation}
where we choose an orthogonal Cartesian coordinate system ($\tilde{\phi}\in [0,2\pi],\, \tilde{\theta}\in [0,2\pi]$) for the C$\nu$B frame and the quantities $q$ in this frame are denoted by $\tilde{q}$. $|\vec{\tilde{p}}_{\nu_i}|$ is the average relic neutrino momentum. In general, one cannot assume that the Earth is stationary with respect to this frame (for instance, the Earth is moving relative to the Cosmic Microwave Background (CMB) which gives rise to the CMB dipole \cite{Hoffman_2015,Planck18}). If the Earth is moving at a speed $\beta_{\oplus}$ w.r.t the C$\nu$B frame and we orient our coordinate system such that $\hat{\beta}_{\oplus}=\hat{z}$, then one can find the neutrino momentum in the Earth frame as,
\begin{equation}
	\vec{p}_{\nu_i}=|\vec{\tilde{p}}_{\nu_i}|\text{cos}\,\tilde{\phi}\,\text{sin}\,\tilde{\theta}\,\hat{x}+|\vec{\tilde{p}}_{\nu_i}|\text{sin}\,\tilde{\phi}\,\text{sin}\,\tilde{\theta}\,\hat{y}+\gamma_{\oplus}(|\vec{\tilde{p}}_{\nu_i}|\text{cos}\,\tilde{\theta}-\beta_{\oplus}\tilde{E}_{p_{\nu_i}})\,\hat{z}\,,
\end{equation}
where $\gamma_{\oplus}=1/\sqrt{1-{\beta_{\oplus}}^2}$ and $\tilde{E}_{p_{\nu_i}}=\sqrt{|\vec{\tilde{p}}_{\nu_i}|^2+m^2_{\nu_i}}$ are the Lorentz factor and energy of the neutrino in the C$\nu$B frame respectively. Since, we do not know the velocities of the constituent neutrinos of the C$\nu$B \cite{darkstod} we perform flux averages over all possible relic neutrino orientations as outlined in \cite{Bauer_2023}. The flux average of a quantity $A_{\nu_i}$ dependent upon the dynamics of the neutrino background, in the Earth/lab frame is defined as,
\begin{equation}\label{Fluxavg}
	\langle A_{\nu_i} \rangle=\frac{\int(\vec{\beta}_{\nu_i}\cdot \vec{n}_{\oplus}) A_{\nu_i} \,d\tilde{\Omega}}{\int(\vec{\beta}_{\nu_i}\cdot \vec{n}_{\oplus})\,d\tilde{\Omega}}\,,
\end{equation}
where $\vec{\beta}_{\nu_i}=\vec{p}_{\nu_i}/E_{p_{\nu_i}}$ is the average velocity of the neutrino, and $\vec{n}_{\oplus}=(\text{cos}\,\tilde{\phi}\,\text{sin}\,\tilde{\theta}\,\hat{x}+\text{sin}\,\tilde{\phi}\,\text{sin}\,\tilde{\theta}\,\hat{y}+\text{cos}\,\tilde{\theta}\,\hat{z})$ is the normal vector to the Earth in the lab frame up to $\mathcal{O}(\beta_{\oplus})$ which could be oriented in any arbitrary direction. Note that just like $A_{\nu_i}$, here $\vec{\beta}_{\nu_i}$ depends upon the dynamics of the system of many relic neutrinos since it is related to the average neutrino momentum $|\vec{\tilde{p}}_{\nu_i}|$, which on the other hand is also time-dependent due to the expansion of the Universe.  
The role of neutrinos in the early universe was first discussed in \cite{Alpher53} where it was considered that the neutrinos were in thermal equilibrium with other relativistic species of electrons, photons, and mesons. As the Universe expanded, the weak interaction rate $\Gamma_W$ fell below that of the expansion rate of the Universe, characterized by the Hubble parameter $H$. This is when the neutrinos fell out of equilibrium and their abundances were effectively frozen out \cite{Dolgov_2002,LESGOURGUES_2006}.
Applying entropy conservation in an expanding universe and taking into account the relativistic degrees of freedom of fermions one can estimate the present day value of the average relic neutrino momentum \cite{quigg2008},
\begin{equation}
	|\vec{\tilde{p}}_{\nu_i,0}|\simeq 3.15 \, T_{\nu,0}\,,
\end{equation}
where $T_{\nu,0}$ is the temperature of the C$\nu$B today. With a present day neutrino temperature $T_{\nu,0}\simeq 0.168\, \text{meV}$ \cite{quigg2008,Dolgov_2008} one can calculate the average momentum to be $|\vec{\tilde{p}}_{\nu_i,0}|\simeq 0.5292\,\text{meV}$. Fermionic species like neutrinos in thermal equilibrium must follow the Fermi-Dirac distribution. After freeze-out, the neutrinos should follow the same distribution (assuming no further interactions) with the energies redshifted due to the expansion of the Universe \cite{Ferrer_2000}. The present day neutrino number density per degree of freedom can be calculated to be,
\begin{equation}
	n_{\nu,0}=\frac{3\zeta(3)T^3_{\nu,0}}{4\pi^2}\simeq 56\,\text{cm}^{-3}\,.
\end{equation}
Since, only left chiral neutrinos and right chiral anti-neutrinos exist in the SM, we expect that all ultra-relativistic neutrinos (anti-neutrinos) at freeze-out will be left (right)-helicity particles. We consider the relic neutrinos today to be mass eigenstates since they would have long decohered from their flavor superpositions at time scales much shorter than cosmological timescales. Then the abundances for Dirac neutrinos today is,
\begin{subequations}
		\label{AbD}
\begin{align}
	&\tilde{n}_\nu(\nu^D_{i,L})\simeq n_{\nu,0}~~,~~~\tilde{n}_\nu(\nu^D_{i,R})\simeq 0\,, \\
	&\tilde{n}_\nu(\Bar{\nu}^D_{i,L})\simeq 0~~~~,~~~\tilde{n}_\nu(\Bar{\nu}^D_{i,R})\simeq n_{\nu,0}\,,
\end{align}
\end{subequations}
and for Majorana neutrinos,
\begin{align}\label{AbM}
	&\tilde{n}_\nu(\nu^M_{i,L})\simeq n_{\nu,0}~~,~~~\tilde{n}_\nu(\nu^M_{i,R})\simeq n_{\nu,0}\,.
\end{align}
We call this the standard scenario of relic neutrino abundances. Using the aforementioned neutrino number densities, we see that the energy shift due to Majorana neutrinos completely vanishes, see Eq. (\ref{finalEM}), while that for Dirac neutrinos is non-zero only in the presence of tensor parameters  and is independent of the NSI parameters Eq. (\ref{finalE}). We note that in the absence of BSM couplings, the energy shift due to Dirac neutrinos completely vanishes in the SM scenario. This is in disagreement with Eq. (5.34) of \cite{Bauer_2023}, wherein the helicity asymmetry contribution is non-zero for the standard decoupling abundances, see Eq. (\ref{AbD}). We find that the origin of this discrepancy is in a missing negative sign in the helicity asymmetry term of the formula in Ref. \cite{Bauer_2023}. Our result is in agreement with \cite{Duda_2001,Perez-Gonzalez:2024xgb} in the limit of the SM.
Note that the neutrino number densities of different helicities will in general be frame dependent since helicity is not a Lorentz invariant quantity. Thus, we need to find out the abundances in the Earth frame using the corresponding abundances in the C$\nu$B frame. Whether the helicity of a relic neutrino will flip or not depends upon the motion of the Earth through the C$\nu$B frame. For our case, since the Earth is moving in the $z$ direction relative to the relic neutrino frame, the helicity of neutrinos with velocity component in $-\text{ve}\; z$ direction will never flip. But the helicity of neutrinos with velocity component in the $+\text{ve}\; z$ direction can flip depending on whether or not the velocity of Earth is greater in magnitude. If the component of the velocity of the Earth in the direction of motion of a particular neutrino, $\beta_{\oplus}\text{cos}\,\tilde{\theta}> \tilde{\beta}_{\nu_i} $, then going to the Earth frame will change a left/right-handed neutrino to a right/left-handed one.
One can formulate the Earth frame neutrino number density for a given helicity in terms of the number densities of the C$\nu$B as \cite{Bauer_2023},
\begin{align}\label{helicityflip1}
	&n_{\nu}(\nu_{i,L})=\gamma_{\oplus}\left[P_F(\beta_{\oplus})\tilde{n}_\nu(\nu_{i,R})+(1-P_F(\beta_{\oplus}))\tilde{n}_\nu(\nu_{i,L})\right]\,,\\\label{helicityflip2}
	&n_{\nu}(\nu_{i,R})=\gamma_{\oplus}\left[P_F(\beta_{\oplus})\tilde{n}_\nu(\nu_{i,L})+(1-P_F(\beta_{\oplus}))\tilde{n}_\nu(\nu_{i,R})\right]\,,
\end{align}
where 
\[
P_F(\beta_{\oplus}) = 
\begin{cases}
\frac{1}{\pi}\sin^{-1}\Bigl(\frac{\beta_{\oplus}}{\tilde{\beta}_{\nu_i}}\Bigl) & \text{if } \beta_{\oplus} < \tilde{\beta}_{\nu_i}, \\
\frac{1}{2}   & \text{if } \beta_{\oplus} \geq \tilde{\beta}_{\nu_i}, \\
\end{cases}
\]
is the probability of helicity flip of a relic neutrino with given handedness. We explicitly show these Earth frame number densities for a given scenario in Appendix \ref{appendixB}.
Going forward, we need to choose allowed values of the neutrino masses to perform numerical computations. For mass squared differences of $\Delta m^2_{21}=7.49 ^{+0.19}_{-0.19} \times 10^{-5} \text{eV}^2$ and $\Delta m^2_{32}= 2.513^{+0.021}_{-0.019} \times 10^{-3} \text{eV}^2$ (uncertainties corresponding to one standard deviation) in the Normal Hierarchy \cite{Esteban_2024}, we find that $m_2 \gtrsim 8.65\,\text{meV} $ and $m_3  \gtrsim 50.13\,\text{meV} $.
We consider that in the lab/Earth frame our experimental material contains electrons which are at rest ($|\vec{p}_e|=0$) and their spin polarization is along the $z$ axis such that $S^\mu_e=(0,0,0,s_e)$. We also take note that our most general Lagrangian, Eq. \eqref{GNILag}, can also be parametrized in a different way \cite{Rodejohann_2017, BISCHER19} as,
\begin{equation}\label{diffLag}
	\mathcal{L}=\frac{G_F}{\sqrt{2}}\sum_{a=S,P,V,A,T}(\Bar{\nu}_\alpha \Gamma^a \nu_\beta)[\Bar{e}\,\Gamma^a(C^a_{\alpha\beta}+D^a_{\alpha\beta}i^a \gamma^5)e]\,,
\end{equation}
where $\Gamma^a\equiv \{ \mathds{1},i\gamma^5,\gamma^\mu,\gamma^\mu \gamma^5, \sigma^{\mu\nu}=\frac{i}{2}[\gamma^\mu,\gamma^\nu]\}$, corresponding to scalar, pseudoscalar, vector, axial-vector, and tensor operators respectively. One sets $i^a=i$ for $a=(S,P,T)$ and $i^a=1$ for the rest. The inclusion of an imaginary factor in the interaction term renders $C^a_{\alpha\alpha}$ and $D^a_{\alpha\alpha}$ real. The parameters $C^a$ and $D^a$ can be related to the parameters $\epsilon^k$ and $\tilde{\epsilon}^k$ of Eq. (\ref{GNILag}) by simple algebra. The SM is realised in this parametrization by setting all the parameters to zero except the following,
\begin{equation}
	C^{V,SM}=-D^{A,SM}=\epsilon^{L,SM}+\epsilon^{R,SM}~,~~~D^{V,SM}=-C^{A,SM}=-\epsilon^{L,SM}+\epsilon^{R,SM}\,.
\end{equation}
\begin{figure}[t]
    \centering
    \includegraphics[width=0.45\linewidth]{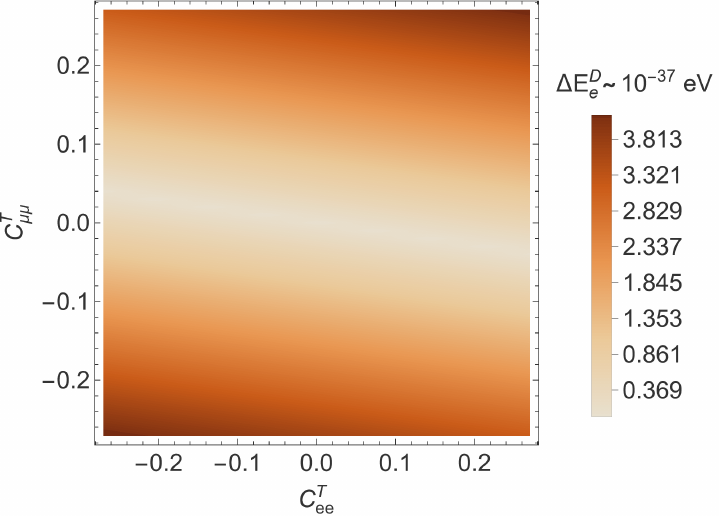}
    \hfill
    \vspace{1em}
    \includegraphics[width=0.45\linewidth]{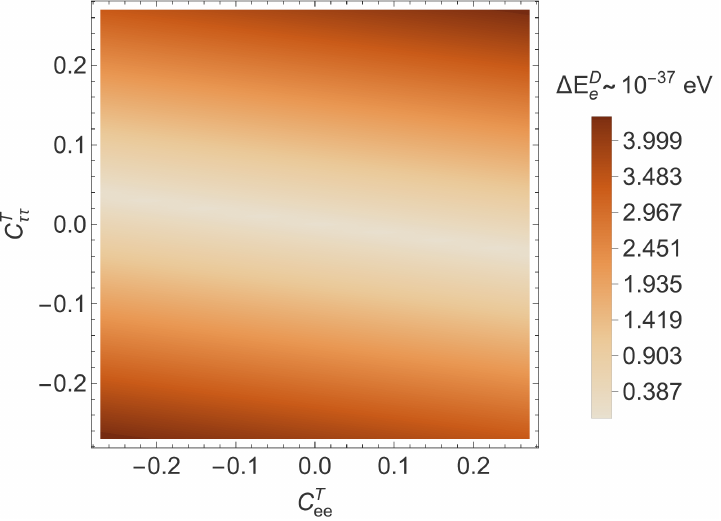}
    \hfill
    \includegraphics[width=0.45\linewidth]{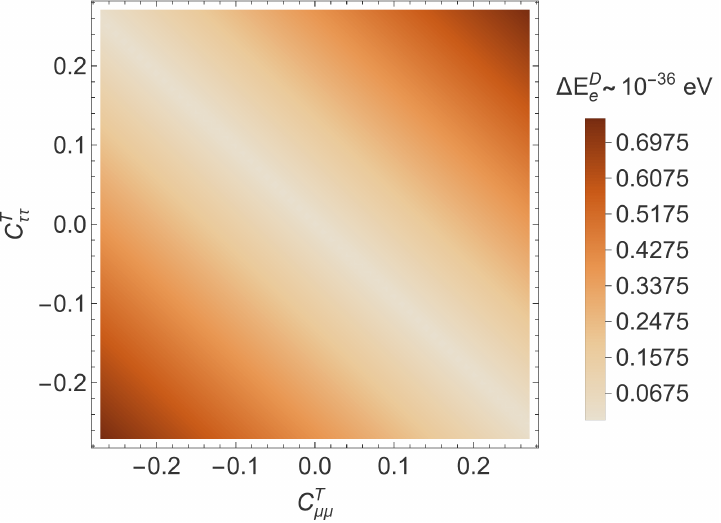}
    \hfill
    \includegraphics[width=0.45\linewidth]{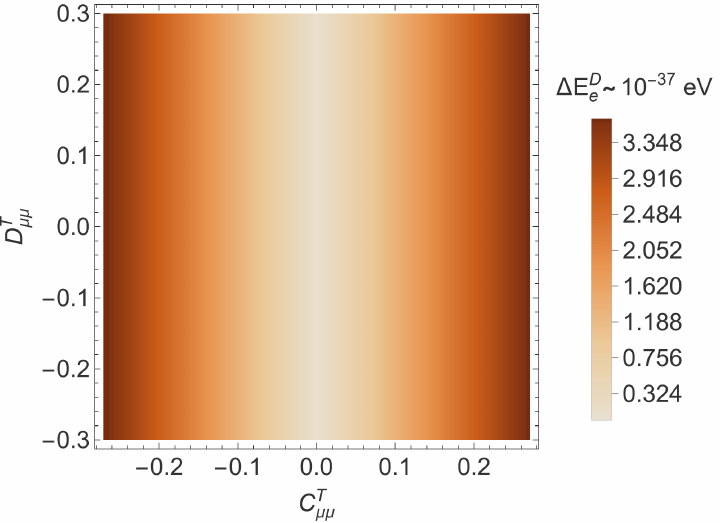}
    \hfill
    \caption{The dependence of the energy splitting is shown with respect to the various tensor parameters in the standard Dirac scenario Eq. (\ref{AbD}) of relic neutrino abundances. In our numerical calculations we have taken $\beta_{\oplus}\simeq 10^{-3}$ which is the case if relic neutrinos are unclustered for which the C$\nu$B frame coincides with CMB \cite{FirstCMB,LucaAmendola_2011}. We use the constraints in \cite{Amir_Rode}, derived only for the diagonal terms and explore the allowed parameter space.}
    \label{fig:1}
\end{figure}
In Fig. \ref{fig:1} we show the dependency of $\Delta E^D_e$ on various diagonal tensor parameters. The NSI contributions cancel out in the standard scenario of relic neutrino number densities, leaving the sole dependence upon the real tensor parameters $C^T$ and $D^T$. These are related to the tensor parameters $\epsilon^T$ and $\tilde{\epsilon}^T$ of our original parametrization Eq. (\ref{GNILag})  as,
\begin{equation}
	\epsilon^T=\frac{1}{4}(C^T-i D^T)~,~~~\tilde{\epsilon}^T=\frac{1}{4}(C^T+i D^T).
\end{equation}
Note that since Eq. (\ref{finalE}) is only dependent upon the sum of the two tensor parameters and not their difference, the energy splitting is only dependent upon $C^T$ and shows no variation with respect to $D^T$. From Fig. \ref{fig:1} we see that the energy shift can reach an order of magnitude of $10^{-36}$ eV within the allowed parameter range. The flux averages associated with the tensorial contribution are significantly enhanced due to the presence of the electron 4-momenta $p_e$. This enhancement is not present for the SM and NSI sectors (see Eq. (\ref{finalE}) and (\ref{finalEmod})). A representative example of this phenomenon is depicted in Appendix \ref{appendixB}.
In Fig. \ref{fig:2} we take a look at how the off-diagonal tensor couplings affect the energy splitting for Dirac neutrinos assuming the standard relic abundances of such neutrinos. The splitting is of the order of $10^{-37}-10^{-36}\,\text{eV}$. Such an energy scale is indeed extremely difficult to probe with current technology. 
Since the energy splitting of a particular electron is so small, we must consider macroscopic objects as relic neutrino detectors.
The strongest commercially available ferromagnets are made of Neodymium alloys \cite{Ndcomm1, Ndcomm2}. The crystalline phase $\text{Nd}_{2}\text{Fe}_{14}\text{B}$ is responsible for the magnetic properties of these materials \cite{Ndstructure}. The bulk magnetic moment of $\sim \text{35}\,\mu_B$ per formula unit \cite{Ndmagnetic} causes the material to have many unpaired electrons aligned in the same direction. The molar mass corresponding to the formula unit is around 1081.12 g\,$\text{mol}^{-1}$, which amounts to $\sim 10^{28}$ unpaired electrons in $10^3$ kg of the material. Their spins can be aligned in our reference $z$ direction.
Following the arguments which led up to Eq. (\ref{ferroacc}), we can take two spherical $\text{Nd}_{2}\text{Fe}_{14}\text{B}$ ferromagnets which are each at a distance $R=1$ cm from some common central axis and align them in such a way that their magnetic moments (direction of spin) are anti-parallel to each other. Then  the resulting acceleration is,
\begin{align}
	a \approx 10^{11}\times|\Delta E_e|(\text{eV})\, \text{cm}\, \text{s}^{-2}=10^{-25} - 10^{-26}\, \text{cm}\, \text{s}^{-2}.
\end{align}
Moreover, torsion balances utilising Meissner suspension consisting of a superconducting body floating above a superconducting coil have been considered in the literature \cite{hagmann1999cosmic} and the expected sensitivity to energy splittings is of the order of $10^{-36}$eV \cite{darkstod}. Recently, levitated ferromagnetic torsional oscillators have been considered for ultralight axion dark matter coupled to electron spins \cite{Li:2026rty,Kalia:2024eml}. Hence, it is possible that such a detector built in the near future will be able to probe dynamical effects arising due to the cosmic neutrino wind after excluding the background of neutrinos from other sources.
\begin{figure}[t]
    \centering
    \includegraphics[width=1.0\linewidth]{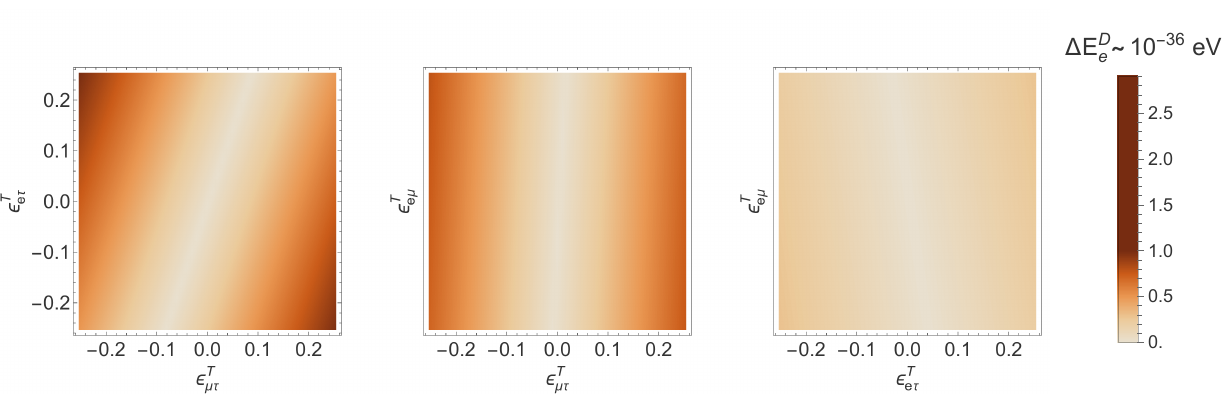}
     \caption{The dependence of the energy splitting is shown with respect to the various off-diagonal tensor parameters $\epsilon^T_{e\tau}, \epsilon^T_{e\mu}$ and $\epsilon^T_{\mu\tau}$. All the parameters are assumed to be real. We use the constraints in \cite{Escrihuela:2021mud}, derived for the off-diagonal terms and explore the allowed parameter space.}
    \label{fig:2}
\end{figure}

\subsection{Asymmetric neutrino background}
\label{subsec:asymm}
It is generally assumed that the C$\nu$B has a vanishingly small asymmetry between neutrino and anti-neutrino number densities. From both theoretical considerations \cite{Dolgov:1991fr} and bounds from Big Bang Nucleosynthesis \cite{DolgovBBN_2002, PhysRevD.66.025015, PhysRevD.66.013008, Ruchayskiy23} one concludes that the asymmetry is negligible for all practical purposes. But even for an originally symmetric C$\nu$B background after decoupling, processes in the Universe yet unknown could have altered the local number densities. BSM physics scenarios often allow for such large asymmetries \cite{Foot_1996, Dev:2024yrg}. Moreover, overdensities produced due to refraction in the interior of the Earth have also been considered \cite{gruzinov2024, huang2024,kalia2024}. In order to try and explain the baryon asymmetry of the Universe (BAU), leptogenesis has been used as a pathway to baryogenesis in the literature \cite{Fukugita:1986hr, KUZMIN198536}. In this scenario, generating a large lepton asymmetry is not possible in the charged sector due to the electrical neutrality of the early universe. But relic neutrinos being neutral leptons allow for such large asymmetries \cite{Borah:2022uos, Mauro23}. 
\begin{figure}[p]
    \centering
    \includegraphics[width=0.45\linewidth]{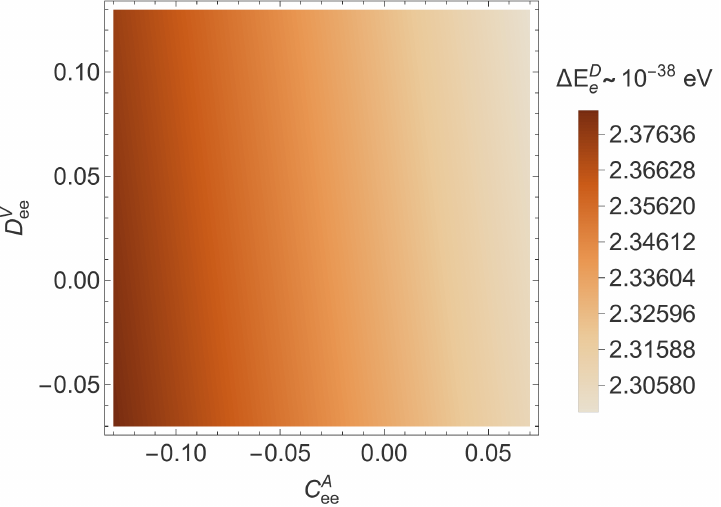}
\hfill
    \includegraphics[width=0.45\linewidth]{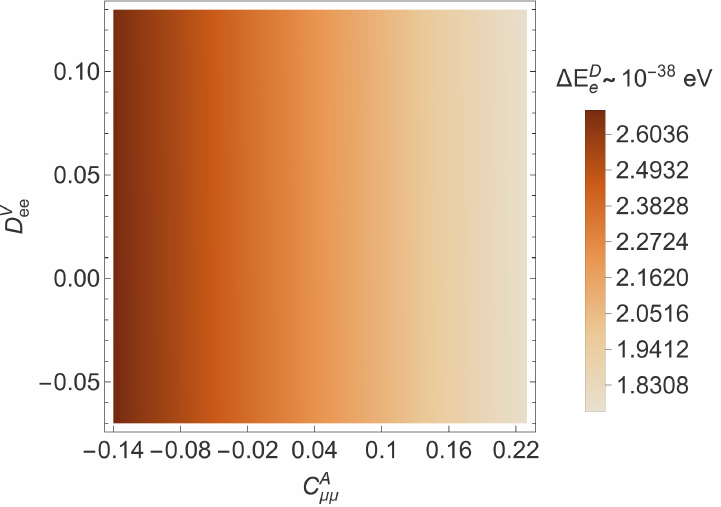}
\hfill
    \includegraphics[width=0.45\linewidth]{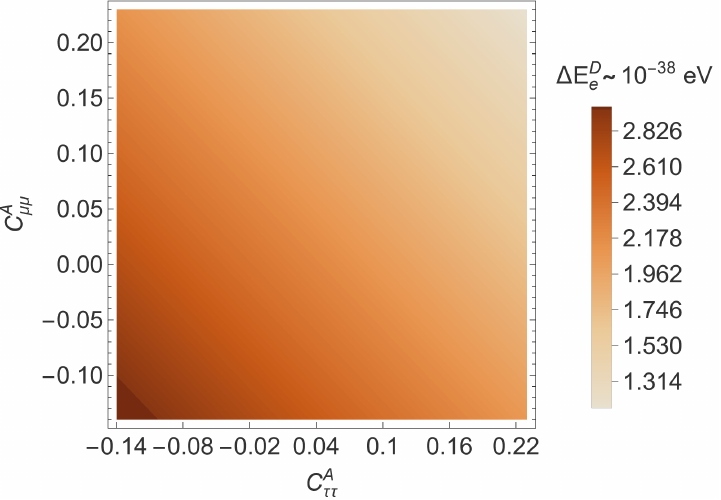}
\hfill
    \includegraphics[width=0.45\linewidth]{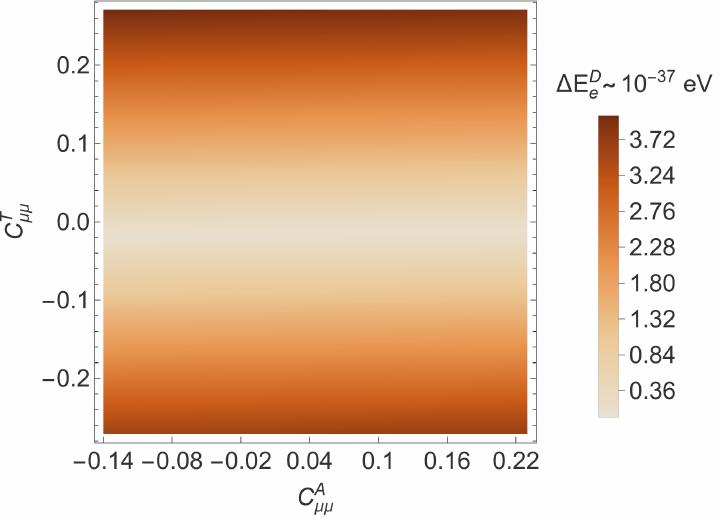}
\hfill
    \includegraphics[width=0.45\linewidth]{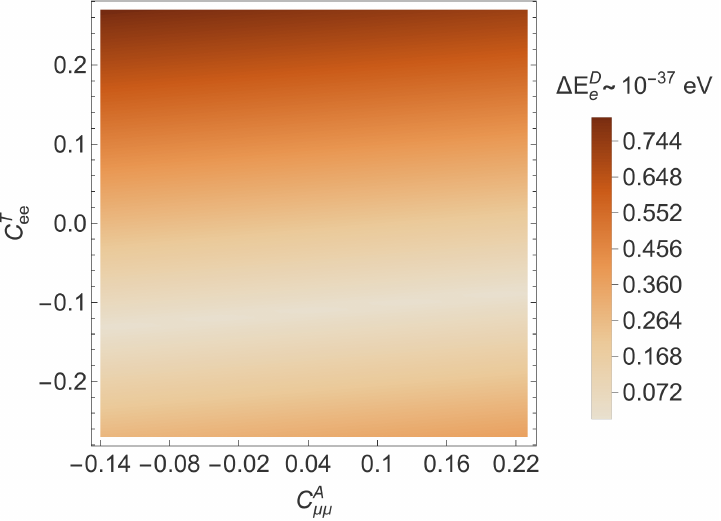}
\hfill
    \includegraphics[width=0.45\linewidth]{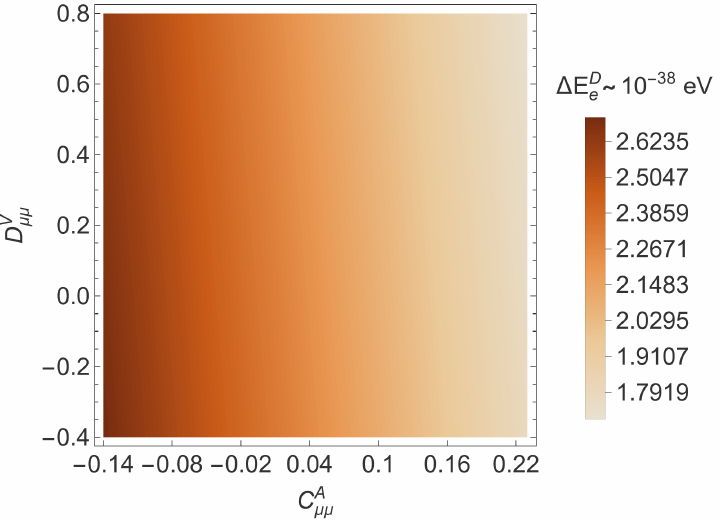}
\hfill
    \includegraphics[width=0.45\linewidth]{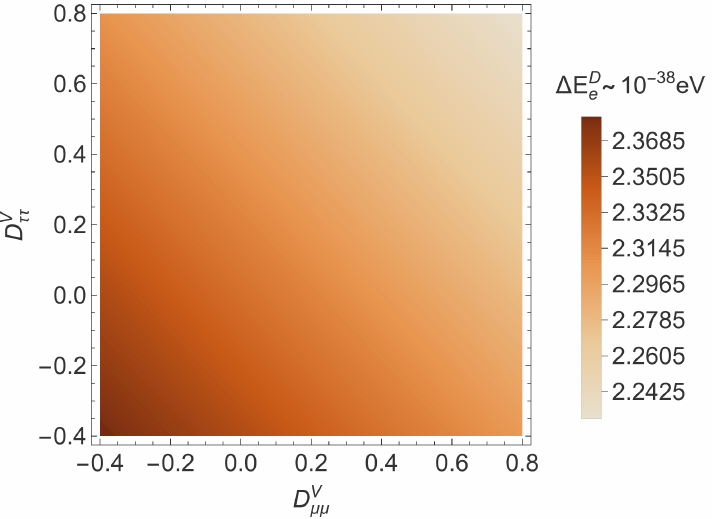}
\hfill
    \includegraphics[width=0.45\linewidth]{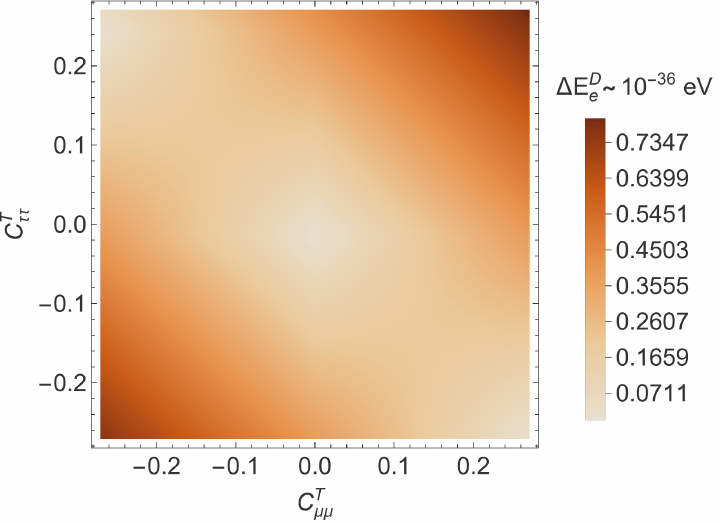}
     \caption{The dependence of the Dirac energy splitting is shown with respect to the various NSI and tensor parameter ranges \cite{Amir_Rode}. For our calculation we set $\eta_\nu = 0.01$ and assume that there is an excess of anti-neutrinos over neutrinos. The non-zero Standard Model contribution is of the order of $10^{-38}$ eV.}
    \label{fig:3}
\end{figure}
Hence, it is of general interest to analyse our formulae for the energy shifts for Dirac and Majorana neutrinos in the ``non-standard" scenario, i.e. we assume deviations from Eq. (\ref{AbD}), such that there is an asymmetry between the neutrino and anti-neutrino number densities for the Dirac case and an asymmetry between left-handed and right-handed number densities for the Majorana case. 
\begin{figure}[t]
    \centering
    \includegraphics[width=1.0\linewidth]{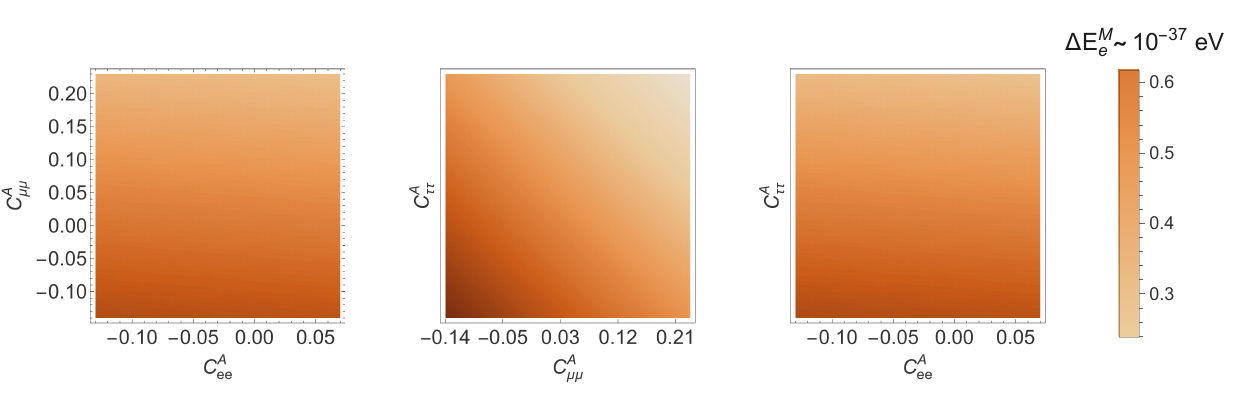}
    \hfill
    \caption{The dependence of the Majorana energy splitting is shown with respect to the contributing NSI parameters \cite{Amir_Rode}. This phenomenon is absent in the case of standard decoupling neutrino abundances. The non-zero Standard Model contribution is of the order of $10^{-38}$ eV.}
    \label{fig:4}
\end{figure}
In Section \ref{sec:Stodolsky} we took note of the fact that the energy shift, shown in Eq. (\ref{finalEM}), for the Majorana case depends only on the NSI parameters. On the other hand, for the Dirac case we saw that the energy shift Eq. (\ref{finalE}) is dependent on both NSI and tensor parameters. We recall that out of all the NSI parameters, $\epsilon^L$ and $\epsilon^R$ have non-zero SM contributions and as such we will look at the variations on top of these. Thus, our calculations for the energy shift will not only have BSM contributions but also have the base SM value about which the fluctuations occur. This SM contribution is absent in the standard Dirac scenario of relic abundances. In order to find out the energy shifts, we make use of the asymmetry parameter $\eta_{\nu}=({n_{\nu}-n_{\bar{\nu}}})/{n_{\gamma}}$, where $n_{\gamma}={2\zeta(3)T^{3}_{\gamma,0}}/{\pi^2}$, with $T_{\gamma,0}$ denoting the present day CMB temperature. We consider $\eta_{\nu}\sim \mathcal{O}(10^{-2}) $ in line with the literature trying to resolve the Helium anomaly which initially arose from the measurements of $^4$He abundance from extremely metal poor galaxies \cite{Mauro23,Kawasaki_2022}. This is in line with the recent bounds set by BBN and CMB observations \cite{Domcke:2025jiy}. 
In a model independent way, we assume that this asymmetry is between the present day neutrino and anti-neutrino number densities for all three mass eigenstate degrees of freedom and find out the energy shift for the Dirac scenario. For the Majorana scenario we take the aforementioned asymmetry between the helicity states.
Mapping Eqs. (\ref{GNILag}) and (\ref{diffLag}) gives us the relations,
\begin{subequations}
\begin{align}
	&\epsilon^L=\frac{1}{4}(C^V-D^V+C^A-D^A)~,~~~\epsilon^R=\frac{1}{4}(C^V+D^V-C^A-D^A) \,,\\
	&\tilde{\epsilon}^L=\frac{1}{4}(C^V-D^V-C^A+D^A)~,~~~\tilde{\epsilon}^R=\frac{1}{4}(C^V+D^V+C^A+D^A)\,,\\
	&\epsilon^T=\frac{1}{4}(C^T-i D^T)~,~~~\tilde{\epsilon}^T=\frac{1}{4}(C^T+i D^T)\, ,
\end{align}
\end{subequations}

and thus we can write our energy shift as,

\begin{align}\label{finalEmod}
	\Delta E^D_e (\vec{0},s_e)&=\frac{G_F}{\sqrt{2}}\sum_{i,s}\Biggr[\Bigl(n_{\nu}(\nu^D_{i,s})-n_\nu ({\Bar{\nu}}^D_{i,s})\Bigl)\Biggl\{D^V_{ii}\biggl\langle\frac{S_e \cdot p_{\nu_i}}{E_{p_{\nu_i}}} \biggl\rangle+\frac{2}{m_e}C^T_{ii}\biggl(\biggl\langle\frac{(S_{\nu_i}\cdot p_e)(S_e \cdot p_{\nu_i})}{E_{p_{\nu_i}}} \biggl\rangle\notag\\ &-\biggl\langle\frac{(p_e \cdot p_{\nu_i})(S_e \cdot S_{\nu_i})}{E_{p_{\nu_i}}} \biggl\rangle\biggl) \Biggl\}+\Bigl(n_{\nu}(\nu^D_{i,s})+n_\nu ({\Bar{\nu}}^D_{i,s})\Bigl)\Biggl\{C^A_{ii}m_{\nu_i}
\biggl\langle\frac{S_e \cdot S_{\nu_i}}{E_{p_{\nu_i}}} \biggl\rangle \Biggl\}\Biggr]\notag\\
&+\text{terms independent of}\:{S_e}\,,
\end{align}

\begin{align}\label{finalEMmod}
	\Delta E^M_e (\vec{0},s_e)&={\sqrt{2}}{G_F}\sum_{i,s}\Biggr[n_{\nu}(\nu^D_{i,s})\Biggl\{C^A_{ii}m_{\nu_i}\biggl\langle\frac{S_e \cdot S_{\nu_i}}{E_{p_{\nu_i}}} \biggl\rangle \Biggl\}\Biggr]+\text{terms independent of}\:{S_e}\,.
\end{align}
In Fig. \ref{fig:3} we plot the energy splitting $\Delta E_e$ with respect to the different NSI and tensor parameters considering an asymmetry between neutrino and anti-neutrino number densities. In the case of Majorana neutrinos, we consider an asymmetry between left and right-handed neutrinos instead, such that there is an explicit dependence of the energy shift on the NSI parameters. In Fig. \ref{fig:4} we depict how the energy splitting in the Majorana case varies with different values of the chosen parameters.
From Fig. \ref{fig:3} and Fig. \ref{fig:4} we see that the maximum possible energy splitting for two non-zero parameters at a time is in the range of $10^{-38}\sim10^{-36}$ eV. Thus, probing a particular parameter space can increase the energy splittings and consequently can be more appropriate from the analysis point of view. In contrast with Fig. \ref{fig:1} and Fig. \ref{fig:2} where we dealt with the symmetric standard scenario Eq. (\ref{AbD}), we note that the energy splitting for the asymmetric case has a non-zero value at the origin of each plot. This indicates the fact that in an asymmetric relic neutrino background we have a non-zero contribution from the SM parameters as well.

\subsection{A brief look at flavor eigenstates}
\label{subsec:accreac}
Investigating what happens for flavor eigenstate neutrinos is crucial for solar neutrino detection experiments as well as for terrestrial neutrinos from reactors and accelerators like DUNE. In Section \ref{sec:Stodolsky}, we derived the energy splitting for the Stodolsky effect for mass eigenstate neutrinos interacting with matter. This was performed so that we could consistently analyse the C$\nu$B as it should constitute of neutrinos that have long decohered since decoupling and exist today as mass eigenstates.
In contrast to relic neutrinos, the neutrinos of terrestrial and astrophysical origin can be treated as flavor eigenstates and thus one would have to work in the flavor basis to find out the electronic energy shifts of the material constituting terrestrial test objects. For high energy neutrinos one need not consider a spread out momentum wave-packet and can directly write the neutrino state as,
\begin{align}
	&|p_{\nu_\alpha},s\rangle=\sqrt{2E_{\nu_\alpha}}f^\dagger_{\nu_\alpha}(\vec{p}_{\nu_\alpha},s)|0\rangle\,, \\
	&|p_{\Bar{\nu}_\alpha},s\rangle=\sqrt{2E_{\nu_\alpha}}g^\dagger_{\nu_\alpha}(\vec{p}_{\nu_\alpha},s)|0\rangle\,,
\end{align}
where we use $f^\dagger_{\nu_\alpha}$ and $g^\dagger_{\nu_\alpha}$ to denote the creation operators for neutrinos and anti-neutrinos of a given flavor $\alpha$.
Recalling the GNI Lagrangian, Eq.~(\ref{GNILag}), we have the energy splitting as,
\begin{equation}\label{Energyshiftflav1}
	\Delta E_e(\vec{p}_e,s_e,\vec{p}_{\nu_\alpha})=\sum_{\nu,\alpha,s}\sum_{N_\nu(p_{\nu_\alpha})} \langle e_{p_e,s_e},\nu_{p_{\nu_\alpha},s}|:\int d^3x\, \mathcal{H}(x):|e_{p_e,s_e},\nu_{p_{\nu_\alpha},s}\rangle \,,
\end{equation}
where we now have a dependency on the 3-momenta of the incoming neutrino and $N_{\nu}$ denotes sum over all background neutrinos of a particular kind ($\nu, \alpha, s$) having 3-momentum $\vec{p}_{\nu_\alpha}$.
To work out this expression, we can equivalently consider the momentum distribution shown in Eq.~(\ref{momdistri}) as a delta-function peaked at the particular momentum $\vec{p}_{\nu_\alpha}$.

We note that the neutrino beam contains neutrinos spread over a range of energies. One can in principle, estimate the total neutrino flux as a function of the beam energy $\Phi_\nu(E_\nu)$ or the differential neutrino flux ${d\Phi_\nu}/{d E_\nu}$ as a function of the neutrino energy. Then integrating over the entire energy range results in the prediction for the mean energy shift of a particular electron. Assuming that the beam neutrinos are travelling close to the speed of light, we can approximate $n_\nu \simeq \Phi_\nu$ (in natural units). Then for the former case we can write down our average energy shift as,

\begin{align}\label{AvgEshift}
	\Delta E^{\text{avg}}_e (\vec{p}_e,s_e)
&=\frac{3}{16\pi p^3_f\, E_{p_e} V}\sum_{\nu,\alpha,s}\int d^3 p_{\nu_\alpha}\Phi_{\nu}(\nu_{\alpha,s,p_{\nu_\alpha}}){\frac{{\langle \mathcal{H} \rangle}_\alpha}{E_{p_{\nu_\alpha}}}}\,,
\end{align}

where $p_f$ is the momentum of the highest energy neutrino that the accelerator can produce. The momentum space integral in the denominator is the normalisation factor.
For the differential neutrino flux one has to integrate the energy shift due to $d\Phi$ neutrino flux for neutrino energies ranging from $E_{\nu}$ to $E_{\nu}+dE_{\nu}$. Hence, we have an integrated energy shift of the form,
\begin{align}\label{IntEshiftdiff}
	\Delta E^{\text{int}}_e (\vec{p}_e,s_e)&=\frac{1}{4E_{p_e} V }\sum_{\nu,\alpha,s}\int d E_{p_{\nu_\alpha}}{\frac{{\langle \mathcal{H} \rangle}_\alpha}{E_{p_{\nu_\alpha}}}\biggl(\frac{d\Phi_{\nu}}{dE_\nu}\biggl)_{\alpha,s,p_{\nu_\alpha}}}\,.
\end{align}
Thus, it is possible to calculate the energy shift of the electron making use of either Eq.~(\ref{AvgEshift}) or Eq.~(\ref{IntEshiftdiff}) depending on the type of data available on the incident neutrinos in question. However, since the flux of neutrinos from astrophysical and terrestrial sources is much smaller than that of the C$\nu$B, it is expected that the corresponding energy splittings will be much smaller than the typical values derived in Section \ref{subsec:CnuB} and \ref{subsec:asymm}. Therefore, terrestrial neutrino sources cannot be used as a potential probe of the Stodolsky effect in the near future.

\section{Summary and Conclusion}
\label{sec:sumandconc}
We have shown that the Stodolsky effect can be parametrised in the most general model independent setting using the generalised neutrino interactions. Our results depict that the energy shift of a single electron in a particular spin-state due to an ambient neutrino background or a neutrino beam depends on non-standard neutrino interactions along with tensor interactions on top of the well-established Standard Model weak interaction. 
We have explicitly shown that scalar and pseudoscalar contributions are absent from the expression of the energy shifts as they are not dependent on the spin of the electron. While the case of Dirac neutrinos permits both NSI and tensor parameters, the Majorana scenario gets contributions only from the NSI interactions.
In our discussion on the C$\nu$B we find that the Stodolsky effect is vanishing in the pure SM sector when considering the standard decoupling abundances of neutrino species. This corresponds to the same average present-day number densities for left-handed neutrinos and right-handed anti-neutrinos corresponding to the Dirac scenario. For the Majorana scenario we have the same average number densities for left and right-handed neutrinos. Although the Majorana energy shift remains zero even after considering the most general interaction Lagrangian, interestingly the Dirac energy shift gets a measurable contribution from the tensor parameters. In a real experimental setting, this energy shift will depend upon the precise values of the $\epsilon^T$ parameters which are currently very weakly constrained. For our analysis we make use of such constraints available in the literature and plot how the energy shifts vary while exploring the parameter space, considering two non-zero parameters at a time. We then discuss the possibility of measuring the energy shifts using a torsion balance consisting of a test material of high electron polarisability such as $\text{Nd}_{2}\text{Fe}_{14}\text{B}$.
We also take a look at how the energy shifts behave if there is an asymmetry between the C$\nu$B number densities. In our discussion we pointed out why such a case is of interest from the cosmological point of view and we explicitly depict the energy shifts for various NSI and tensor parameters. We then compare how these parameters, if non-zero can modify the SM energy shifts. Although the coupling strengths for the tensor operators are much smaller in comparison to that of the SM couplings (ensuring the validity of  Eq.~(\ref{GNILag})), since the tensorial contribution is directly proportional to the neutrino number density the resultant energy shifts are enhanced. Once we include all the flux averages and explore the allowed parameter space for the tensor couplings, an enhancement of one or two orders of magnitude is obtained.

For the sake of a complete analysis, we discuss how to approach the Stodolsky effect if our concerned neutrino flux contains neutrino flavour eigenstates and derive the energy shifts which can be explicitly evaluated from the knowledge of the flux distribution of neutrinos.
Our results indicate that considering generalised neutrino interactions can enhance the Stodolsky effect in the context of the C$\nu$B.  The modeling of cosmic neutrinos using quantum kinetic equations and taking into account gravitational clustering is essential for advancing our understanding of the effect of uncertainties in the C$\nu$B on the energy shifts and their sensitivity to BSM physics. As such the development of newer and more sophisticated experimental methods for the detection of the C$\nu$B is a sector which seems promising.

\acknowledgments
SB acknowledges the valuable discussions with Poonam Mehta, Sabila Parveen, and Navaneeth Poonthottathil. SB especially thanks Jack D. Shergold for helpful correspondence. SB and UKD acknowledge support from the Anusandhan National Research Foundation (ANRF), Government of India, under Grant Reference No. CRG/2023/003769.

\appendix
\section{Details of the calculations of the energy shift}\label{appendixA}
This section is dedicated to the analytical estimation of the energy shift. We will largely use the notation and convention adopted in~\cite{Bauer_2023}. 
The external one particle states are, in general, wave-packets with momentum distribution function $\omega(\vec{p},\vec{q})$ peaked around $\vec{p}$ such that~\cite{Smirnov_2022,Blas_2023},
\begin{equation}
	|\psi_{p,s}(\vec{x}_\psi)\rangle=\int \frac{d^3q}{(2\pi)^3\sqrt{2E_{q}}}\omega_{\psi}(\vec{p},\vec{q})e^{-i\vec{q}\cdot \vec{x}_{\psi}}{|q,s\rangle}_{\psi}~~~~~ \text{for}~ \psi=e,\nu\,,
\end{equation}
and the relativistically normalised external one particle states of definite momentum are defined as,
\begin{subequations}
\begin{align}
	&{|p_e,s_e\rangle}=\sqrt{2E_e}f^\dagger_e(\vec{p}_e,s_e)|0\rangle, \\
	&|p_{\nu_i},s\rangle=\sqrt{2E_{\nu_i}}f^\dagger_{\nu_i}(\vec{p}_{\nu_i},s)|0\rangle, \\
	&|p_{\Bar{\nu}_i},s\rangle=\sqrt{2E_{\nu_i}}g^\dagger_{\nu_i}(\vec{p}_{\nu_i},s)|0\rangle,
\end{align}
\end{subequations}
where $f^\dagger$, $f$ and $g^\dagger$, $g$ are the creation and annihilation operators for particles and anti-particles respectively for a given fermionic species.
Performing a spatial average over the region of space $V$ much larger than the distribution of each wave-packet~\cite{Ghosh_2023},
\begin{equation}
	\Delta E_e(\vec{p}_e,s_e) \rightarrow \frac{1}{V^2}\int d^3 x_{\nu_i} d^3 x_{e}\Delta E_e(\vec{p}_e,s_e)\,,
\end{equation}
we can write the energy shift as,
\begin{align}
	\Delta &E_e(\vec{p}_e,s_e)=\sum_{\nu,i,s}\sum_{N_\nu} \frac{1}{V^2}\int d\Pi\: \omega_{\nu_i}(\vec{p}_{\nu_i},\vec{q}_{\nu_i})\omega^*_{\nu_i}(\vec{p}_{\nu_i},\vec{q}^{\:\prime}_{\nu_i})\omega_{e}(\vec{p}_{e},\vec{q}_{e})\omega^*_{e}(\vec{p}_{e},\vec{q}^{\:\prime}_{e})\notag\\
	& \times \int d^3x_{\nu_i}d^3x_e e^{i({\vec{q}^{\,\prime}_{\nu_i}}-\vec{q}_{\nu_i})\cdot\vec{x}_{\nu_i}} e^{i({\vec{q}^{\,\prime}_{e}}-\vec{q}_{e})\cdot\vec{x}_{e}} \langle q_e,s_e|\otimes\langle q_{\nu_i},s|:\int d^3x\mathcal{H}(x):|q_{\nu_i},s\rangle \otimes|q_e,s_e\rangle \,,
\end{align}
where we define
\begin{equation}
	d\Pi=\frac{d^3 q_{\nu_i} d^3 q^{\prime}_{\nu_i}d^3 q_{e} d^3 q^{\prime}_{e}}{(2\pi)^{12}\,4 \sqrt{E_{q_{\nu_i}}E_{q^{\prime}_{\nu_i}}E_{q_e} E_{q^{\prime}_e}}}\,.
\end{equation}
Carrying out the integration with respect to the spatial coordinates of each of the particles we get
\begin{align}
	\Delta E_e(\vec{p}_e,s_e)=\sum_{\nu,i,s}\sum_{N_\nu} &\frac{1}{V^2}\int\frac{ d^3 q_{\nu_i}d^3q_e}{(2\pi)^6 4E_{q_{\nu_i}} E_{q_e}} |\omega_{\nu_i}(\vec{p}_{\nu_i},\vec{q}_{\nu_i})|^2|\omega_{e}(\vec{p}_{e},\vec{q}_{e})|^2 \langle \mathcal{H}\rangle\,,
\end{align}
where,
\begin{equation}
	\label{Avgnot}
	{\langle \mathcal{H}\rangle}_i=\langle q_e,s_e|\otimes\langle q_{\nu_i},s|:\int d^3x\,\mathcal{H}(x):|q_{\nu_i},s\rangle \otimes|q_e,s_e\rangle \,.
\end{equation}
Noting that $\omega_{\psi}$ is the momentum wavepacket of the corresponding fermionic particle and that $\int {d^3 q}\,|\omega_{\psi}(p,q)|^2={(2\pi)^3}$; we can write~\cite{darkstod},
\begin{equation}\label{momdistri}
	\sum_{N_\nu}\frac{|\omega_{\nu_i}(p_{\nu_i},q_{\nu_i})|^2}{V}=n_\nu(\nu_{i,s})f_{\nu_i}(\vec{q}_{\nu_i}),~~~\frac{|\omega_{e}(p_{e},q_{e})|^2}{V}=\frac{(2\pi)^3}{V} \delta^{(3)}(\vec{p}_e - \vec{q}_e)\,,
\end{equation}
where $n_{\nu}(\nu_{i,s})$ is the number density of neutrinos of a given mass and helicity and $f_{\nu_i}(\vec{q}_{\nu_i})$ is the momentum distribution of the system of neutrinos in the region $V$.
Then the energy shift for an electron of given spin becomes,
\begin{align}
	\label{Energyshift2}
	\Delta E_e (\vec{p}_e,s_e)&=\sum_{\nu,i,s}\frac{n_{\nu}(\nu_{i,s})}{4E_{p_e} V}\int \frac{d^3 q_{\nu_i}}{(2\pi)^3} f_{\nu_i}(\vec{q}_{\nu_i})\frac{{\langle \mathcal{H} \rangle}_i}{E_{q_{\nu_i}}}\, ,\notag\\
	&=\frac{1}{4E_{p_e} V}\sum_{\nu,i,s}n_{\nu}(\nu_{i,s}){\Bigg\langle\frac{{\langle \mathcal{H} \rangle}_i}{E_{q_{\nu_i}}}\Bigg\rangle}.
\end{align}
Here $\langle \langle \mathcal{H} \rangle/ E_{q_{\nu_i}}\rangle$ represents the flux averaging of the quantity $ \langle \mathcal{H} \rangle/ E_{q_{\nu_i}}$. In order to find out $\langle \mathcal{H} \rangle$ we need to write down the concerned fields in terms of the respective creation and annihilation operators. The mode expansion of electron and neutrino Dirac fields is given in the standard form as,
\begin{subequations}
\begin{align}
	&\psi_D(x)=\int \frac{d^3 p}{(2\pi)^3 \sqrt{2 E_p}}\sum_s (f(\vec{p},s)u(p,s)e^{-ip.x}+g^\dagger(\vec{p},s)v(p,s)e^{ip.x})\,,\\
	&\Bar{\psi}_D(x)=\int \frac{d^3 p}{(2\pi)^3 \sqrt{2 E_p}}\sum_s (g(\vec{p},s)\Bar{v}(p,s)e^{-ip.x}+f^\dagger(\vec{p},s)\Bar{u}(p,s)e^{ip.x})\,.
\end{align}
\end{subequations}
While the same for Majorana neutrinos can be given as, $(\psi_M=\mathcal{C}\,\Bar{\psi}^T_M)$,
\begin{align}
	&\psi_M(x)=\int \frac{d^3 p}{(2\pi)^3 \sqrt{2 E_p}}\sum_s (a(\vec{p},s)u(p,s)e^{-ip.x}+a^\dagger(\vec{p},s) \mathcal{C}\Bar{u}^T(p,s)e^{ip.x})\,.
\end{align}
For our calculation, we will consider the general interaction lagrangian presented in Eq.~(\ref{GNILag}) containing all possible Lorentz invariant structures. Using the PMNS matrix~\cite{GRIBOV1969493, PMNS} we switch from the neutrino flavor basis to the mass basis such that the interaction Hamiltonian has the form,
\begin{equation}\label{GNILagMass}
	\mathcal{H}_D=\frac{G_F}{\sqrt{2}}\sum_{\alpha,\beta,j,l} \sum^{10}_{k=1}\: {\stackrel{(\sim)k}{\epsilon_{\alpha \beta}}}\:U^*_{\alpha j}U_{\beta l}(\Bar{\nu}_j\mathcal{O}_k \nu_l)\,(\Bar{e}\mathcal{O}^\prime_k e)\,,
\end{equation}
where $U_{\alpha j}$ denotes the components of the PMNS matrix. We can then calculate that for Dirac neutrinos and anti-neutrinos the averaged out interaction Hamiltonian is of the form (using the notation of Eq.~(\ref{Avgnot})),
\begin{subequations}
\begin{align} 
	\label{Diracexp1}
	&{\langle \mathcal{H}_D\rangle}_i=\frac{G_F}{\sqrt{2}}V\sum^{10}_{k=1}[\Bar{u}(p_{\nu_i},s)\mathcal{O}_k u({p_{\nu_i},s})]\,j^{k}_i\,,\\ \label{Diracexp2}
	&{\langle \Bar{\mathcal{H}}_D\rangle}_i=-\frac{G_F}{\sqrt{2}}V\sum^{10}_{k=1}[\Bar{v}(p_{\nu_i},s)\mathcal{O}_k v({p_{\nu_i},s})]\,j^{k}_i  \,,
\end{align}
\end{subequations}
where,
\begin{equation}
	j^k_i={\stackrel{(\sim)k}{\epsilon_{\alpha \beta}}}\:U^*_{\alpha i}U_{\beta i} \,\Bar{u}(p_e,s_e)\mathcal{O}^\prime_k u(p_e,s_e)\,,
\end{equation}
acts as an electron current which can be a scalar, vector or tensor current for different values of $k$ in Table \ref{table:Tab1}.
The Majorana counterpart of Eq.~(\ref{GNILagMass}) is,
\begin{equation}\label{GNILagMajorana}
	\mathcal{H}_M=\frac{G_F }{\sqrt{2}}\sum_{\alpha,\beta} \sum^{10}_{k=1} {\stackrel{(\sim)k}{\epsilon_{\alpha \beta}}} U^*_{\alpha j}U_{\beta l}(-{\nu}^T_j \mathcal{C}^\dagger \mathcal{O}_k \nu_l)\,(\Bar{e}\mathcal{O}^\prime_k e)\,,
\end{equation}
with additional constraints imposed on the epsilon parameters according to Eq.~(\ref{MajoranaConstraints}). Additionally, we can calculate that
\begin{align}
	\label{HMaj}
	{\langle \mathcal{H}_M\rangle}_i=\frac{G_F}{\sqrt{2}}V\sum^{10}_{k=1}[\Bar{u}(p_{\nu_i},s)\{\mathcal{C}\mathcal{O}^T_k \mathcal{C}^\dagger+\mathcal{O}_k\} u({p_{\nu_i},s})]j^{k}_i\,.
\end{align}
%
Using the properties of the unitary charge conjugation matrix $\mathcal{C}$, it can be shown that the vector and all the tensor operators cancel out in Eq.~(\ref{HMaj}) and only the pseudoscalar ($\gamma_5$) and pseudovector ($\gamma^\mu\gamma_5$) operators survive (since $\mathcal{C}=\mathcal{C}^\dagger$,  $\mathds{1}$ survives as well). The Eqs.~(\ref{Diracexp1}, \ref{Diracexp2}) and (\ref{HMaj}) can also be written in terms of traces as,
\begin{subequations}
\begin{align} 
	\label{DiracTr1}
	&{\langle\mathcal{H}_D\rangle}_i=\frac{G_F}{\sqrt{2}}V\sum^{10}_{k=1}\text{Tr}\Bigr[u({p_{\nu_i},s})\Bar{u}(p_{\nu_i},s)\mathcal{O}_k \Bigr] \text{Tr}\Bigr[{\stackrel{(\sim)k}{\epsilon_{\alpha \beta}}}\:U^*_{\alpha i}U_{\beta i}\,u(p_e,s_e) \Bar{u}(p_e,s_e)\mathcal{O}^\prime_k \Bigr]\,, \\ \label{DiracTr2}
	&{\langle \Bar{\mathcal{H}}_D\rangle}_i=-\frac{G_F}{\sqrt{2}}V\sum^{10}_{k=1}\text{Tr}\Bigr[v({p_{\nu_i},s})\Bar{v}(p_{\nu_i},s)\mathcal{O}_k \Bigr] \text{Tr}\Bigr[{\stackrel{(\sim)k}{\epsilon_{\alpha \beta}}}\:U^*_{\alpha i}U_{\beta i}\,u(p_e,s_e) \Bar{u}(p_e,s_e)\mathcal{O}^\prime_k \Bigr]\,,\\ \label{MajTr}
	&{\langle \mathcal{H}_M\rangle}_i=\frac{G_F }{\sqrt{2}}V\sum^{10}_{k=1}\text{Tr}\Bigr[u({p_{\nu_i},s})\Bar{u}(p_{\nu_i},s)\{\mathcal{C}\mathcal{O}^T_k \mathcal{C}^\dagger+\mathcal{O}_k\}\Bigr] \text{Tr}\Bigr[{\stackrel{(\sim)k}{\epsilon_{\alpha \beta}}}\:U^*_{\alpha i}U_{\beta i}\,u(p_e,s_e) \Bar{u}(p_e,s_e)\mathcal{O}^\prime_k \Bigr]\,. 
\end{align}
\end{subequations}
%
These traces can be evaluated using the $u$ and $v$-spinor identities,
\begin{align} 
	\label{SId1}
	u(p,s)\Bar{u}(p,s)=\frac{1}{2}(\slashed{p}+m)(\mathds{1}+\gamma^5 \slashed{S}) ,~~~v(p,s)\Bar{v}(p,s)=\frac{1}{2}(\slashed{p}-m)(\mathds{1}+\gamma^5 \slashed{S})\,,
\end{align}
where $m$ denotes the mass of the particle and the spin 4-vector for massive particles is defined as,
\begin{equation}
	\label{Spin4}
S^\mu=s\bigg(\frac{\vec{p}\cdot \vec{n}}{m},\vec{n}+\frac{(\vec{p}\cdot \vec{n})\vec{p}}{m(E+m)}\bigg)\,,
\end{equation}
where $s$ is the spin of the particle, $\vec{n}$ is the direction of the spin 3-vector in the rest frame of the particle, and $\vec{p}$ and $E$ are the momentum and energy, respectively. If $\vec{n}={\vec{p}}/{|\vec{p}|}$, then we concern ourselves with the projection of spin onto the direction of the momentum vector and thus, we can identify $s$ with the helicity of the particle. With this, Eq.~(\ref{Spin4}) simplifies to,
\begin{equation}
	\label{Hel4}
S^\mu=s\bigg(\frac{|\vec{p}|}{m},\frac{E \vec{p}}{m|\vec{p}|}\bigg)\,,
\end{equation}
where $s=+1$ (right handed) or $s=-1$ (left handed). One crucial thing to note is that we adopt the same helicity convention for neutrinos and anti-neutrinos throughout this paper to avoid confusion. One can calculate the traces of each of the independent $\gamma$-matrix structures (see Table \ref{table:Tab2}) and find out the final form of the averaged Dirac hamiltonian as,
\begin{table}
\renewcommand{\arraystretch}{1.2}
\centering
\begin{tabular}{l c c l}
\hline
$j$~~  & $\Gamma_j$ & $\text{Tr}[u(p,s)\Bar{u}(p,s)\Gamma_j]$& $\text{Tr}[v(p,s)\Bar{v}(p,s)\Gamma_j]$  \\
\hline
1&  $\mathds{1}$ & ~~$2m$ &~~~~ $-2m$\\
2&  $\gamma^5$ & $-2p^\mu S_\mu$ & ~~~~$-2p^\mu S_\mu$\\
3&  $\gamma^\mu$ & ~~$2p^\mu$ & ~~~~~~~$2p^\mu$\\
4&  $\gamma^\mu \gamma^5$ & ~~$2mS^\mu$ & ~~~~$-2mS^\mu$\\
5&  $\sigma_{\mu\nu}$ & $-2\epsilon_{\alpha\beta\mu\nu}p^\alpha S^\beta$ &~$-2\epsilon_{\alpha\beta\mu\nu}p^\alpha S^\beta$\\
6&  $\sigma_{\mu\nu}\gamma^5$ &~~$2ip_{[\mu} S_{\nu]}$ &~~~~~$2ip_{[\mu} S_{\nu]}$\\
\hline
\end{tabular}
\caption{The $u$ and $v$ traces derived for the different gamma matrix structures. The anti-symmetrization is such that $p_{[\mu} S_{\nu]}=p_\mu S_\nu - p_\nu S_\mu$.}
\label{table:Tab2}
\end{table}
\begin{align}
	\label{avgHD}
{\langle\mathcal{H_D}\rangle}_i&=2\sqrt{2}{G_F V}\bigg[\eta_{\mu\nu}\bigg\{(p^\mu_{\nu_i}-m_{\nu_i}S^\mu_{\nu_i})(p^\nu_{e}-m_{e}S^\nu_{e})(\epsilon^L_{\alpha\beta}U_{\beta i}U^*_{\alpha i})\notag\\
&+(p^\mu_{\nu_i}+m_{\nu_i}S^\mu_{\nu_i})(p^\nu_{e}-m_{e}S^\nu_{e})(\tilde{\epsilon}^L_{\alpha\beta}U_{\beta i}U^*_{\alpha i})+(p^\mu_{\nu_i}-m_{\nu_i}S^\mu_{\nu_i})(p^\nu_{e}+m_{e}S^\nu_{e})(\epsilon^R_{\alpha\beta}U_{\beta i}U^*_{\alpha i})\notag\\
&+(p^\mu_{\nu_i}+m_{\nu_i}S^\mu_{\nu_i})(p^\nu_{e}+m_{e}S^\nu_{e})(\tilde{\epsilon}^R_{\alpha\beta}U_{\beta i}U^*_{\alpha i})\bigg\}+(m_{\nu_i}+p_{\nu_i}\cdot S_{\nu_i})m_e(\epsilon^S_{\alpha\beta}U_{\beta i}U^*_{\alpha i})\notag\\
&+(m_{\nu_i}-p_{\nu_i}\cdot S_{\nu_i})m_e(\tilde{\epsilon}^S_{\alpha\beta}U_{\beta i}U^*_{\alpha i})+(m_{\nu_i}+p_{\nu_i} \cdot S_{\nu_i})(p_e \cdot S_e)(\epsilon^P_{\alpha\beta}U_{\beta i}U^*_{\alpha i})\notag\\
&+(m_{\nu_i}-p_{\nu_i}\cdot S_{\nu_i} )(p_e \cdot S_e)(\tilde{\epsilon}^P_{\alpha\beta}U_{\beta i}U^*_{\alpha i})\notag\\
&+\eta^{\mu\theta}\eta^{\nu \omega}(\epsilon_{\alpha\beta\mu\nu}p^\alpha_{\nu_i}S^\beta_{\nu_i}+i\eta_{\mu\alpha}\eta_{\nu\beta} p^{[\alpha}_{\nu_i} S^{\beta]}_{\nu_i} )(\epsilon_{\gamma\delta\theta\omega}p^\gamma_{e}S^\delta_{e}+i\eta_{\theta\gamma}\eta_{\omega\delta} p^{[\gamma}_{e} S^{\delta]}_{e})(\epsilon^T_{\alpha\beta}U_{\beta i}U^*_{\alpha i})\notag\\
&+\eta^{\mu\theta}\eta^{\nu \omega}(\epsilon_{\alpha\beta\mu\nu}p^\alpha_{\nu_i}S^\beta_{\nu_i}-i\eta_{\mu\alpha}\eta_{\nu\beta} p^{[\alpha}_{\nu_i} S^{\beta]}_{\nu_i} )(\epsilon_{\gamma\delta\theta\omega}p^\gamma_{e}S^\delta_{e}-i\eta_{\theta\gamma}\eta_{\omega\delta} p^{[\gamma}_{e} S^{\delta]}_{e})(\tilde{\epsilon}^T_{\alpha\beta}U_{\beta i}U^*_{\alpha i})\biggl]\,.
\end{align}
One can arrive at ${\langle \Bar{\mathcal{H}}_D \rangle}_i$ from Eq.~(\ref{avgHD}) by replacing $m_{\nu_i}$ with $-m_{\nu_i}$ (see Eq.~(\ref{SId1}))  and putting an overall negative sign upfront. In a similar way, the averaged Majorana Hamiltonian can be given as,
\begin{align}
	\label{avgHM}
{\langle\mathcal{H}_M\rangle}_i&=4\sqrt{2}{G_F V}\bigg[m_{\nu_i} m_e  (S_{e}\cdot S_{\nu_i})(\epsilon^L_{ii}-\tilde{\epsilon}^L_{ii}-\epsilon^R_{ii}+\tilde{\epsilon}^R_{ii})
+(m_{\nu_i}+p_{\nu_i}\cdot S_{\nu_i})m_e\epsilon^S_{ii}\notag\\
&+(m_{\nu_i}-p_{\nu_i}\cdot S_{\nu_i})m_e\tilde{\epsilon}^S_{ii}+(m_{\nu_i}+p_{\nu_i} \cdot S_{\nu_i})(p_e \cdot S_e)\epsilon^P_{ii}+(m_{\nu_i}-p_{\nu_i}\cdot S_{\nu_i} )(p_e \cdot S_e)\tilde{\epsilon}^P_{ii} \biggl]\,,
\end{align}
where we use the notation, $\stackrel{(\sim)k}{\epsilon_{ii}}=\stackrel{(\sim)k}{\epsilon_{\alpha\beta}}U_{\beta i}U^*_{\alpha i}$.

\section{Typical evaluation of $\Delta E_{\text{SM}}/\Delta E_{\text{Tensor}}$}\label{appendixB}
\noindent
The enhancement of $\Delta E_{\text{Tensor}}$ in comparison to $\Delta E_{\text{SM}}$ depends significantly on the flux averages. The numerical value of the flux averages relevant for the tensorial operators are significantly greater than that of the SM. This is due to the fact that the flux averages for the former contains the contribution of the electron 4-momentum $p_e$ which is absent in the latter case. Moreover, the flux averages show significant variation over different neutrino mass eigenstates.

\subsection{$\Delta E_{\text{Tensor}}$}
We take the representative example of the $C^T_{\tau\tau}\, \text{vs.}\, C^T_{\mu\mu}$ plot of Figure \ref{fig:1}. The tensorial energy shift gets its contribution from the part of Eq. (\ref{finalEmod}) depicted in blue,
\begin{align}\label{finalEmodB1}
	\Delta E^D_e (\vec{0},s_e)&=\frac{G_F}{\sqrt{2}}\sum_{i,s}\Biggr[\Bigl(n_{\nu}(\nu^D_{i,s})-n_\nu ({\Bar{\nu}}^D_{i,s})\Bigl)\Biggl\{D^V_{ii}\biggl\langle\frac{S_e \cdot p_{\nu_i}}{E_{p_{\nu_i}}} \biggl\rangle+\textcolor{blue}{\frac{2}{m_e}C^T_{ii}\biggl(\biggl\langle\frac{(S_{\nu_i}\cdot p_e)(S_e \cdot p_{\nu_i})}{E_{p_{\nu_i}}} \biggl\rangle}\notag\\ &\textcolor{blue}{-\biggl\langle\frac{(p_e \cdot p_{\nu_i})(S_e \cdot S_{\nu_i})}{E_{p_{\nu_i}}} \biggl\rangle\biggl)} \Biggl\}+\Bigl(n_{\nu}(\nu^D_{i,s})+n_\nu ({\Bar{\nu}}^D_{i,s})\Bigl)\Biggl\{C^A_{ii}m_{\nu_i}
\biggl\langle\frac{S_e \cdot S_{\nu_i}}{E_{p_{\nu_i}}} \biggl\rangle \Biggl\}\Biggr]\notag\\
&+\text{terms independent of}\:{S_e}\,,
\end{align}
which is directly proportional to $C^T_{ii}$ and can be obtained from the flavour basis coefficients $C^T_{\alpha\beta}$ by a basis transformation using the PMNS matrix: $C^T_{ii}=U^*_{\alpha i}C^T_{\alpha\beta}U_{\beta i}$. From the plot we note that the maximum possible value of energy shift is obtained for the coordinates $(C^T_{\tau\tau},C^T_{\mu\mu})$=($\pm$ 0.271, $\pm$ 0.271). We set our benchmark point (BP) as, 
\begin{equation}
	C^T_{\tau\tau}=0.271\,,~~~~ C^T_{\mu\mu}=0.271\,, ~~~\text{all other}~~ C^T_{\alpha\beta},C^A_{\alpha\beta},D^V_{\alpha\beta}=0\,.
\end{equation}
This choice of BP corresponds to 
%
	$
	C^T=(C^T_{11},C^T_{22},C^T_{33})=(0.088,0.189,0.265)\,.
    $
We now analyse the neutrino number densities in the Earth frame. The C$\nu$B frame abundances for Dirac neutrinos is given by Eq. (\ref{AbD}) of Section \ref{subsec:CnuB}. The non-zero number density value for present day is $n_{\nu,0} \simeq 56\, \text{cm}^{-3}=4.329 \times 10^{-13}\,\text{eV}^3$. Going to the Earth frame (assuming $\beta_{\oplus}\simeq 10^{-3}$) we make use of Eq. (\ref{helicityflip1})-(\ref{helicityflip2}) and find that the number densities (in the units of $10^{-13}\text{eV}^3$) are,
\begin{align}
	&n_\nu(\nu^D_{1,L})= n_\nu(\bar{\nu}^D_{1,R})\simeq 4.328\,,~n_\nu(\nu^D_{2,L})= n_\nu(\bar{\nu}^D_{2,R})\simeq 4.306\,,~ n_\nu(\nu^D_{3,L})= n_\nu(\bar{\nu}^D_{3,R})\simeq 4.198\,, \notag\\
	&n_\nu(\nu^D_{1,R})= n_\nu(\bar{\nu}^D_{1,L})\simeq 1.378\times 10^{-3}\,,~ n_\nu(\nu^D_{2,R})= n_\nu(\bar{\nu}^D_{2,L})\simeq 2.257\times 10^{-2}\,,\notag\\
& n_\nu(\nu^D_{3,R})= n_\nu(\bar{\nu}^D_{3,L})\simeq 1.307\times 10^{-1}\,,
\end{align}
where we see that the motion of the Earth through the C$\nu$B can generate non-zero $\nu_R$ an $\bar{\nu}_L$ densities (although very small).

Next we discuss the role of flux averages in modifying the energy shifts. For the tensorial contribution the flux averages include $p_e$, $p_{\nu_i}$, $S_e$ and $S_{\nu_i}$. For an electron at rest in the Earth frame with spin polarization along the $z$ axis we have $p_e=m_e(1,0,0,0)$ and $S_e=s_e(0,0,0,1)$, where $m_e=0.511 \times 10^6 \,\text{eV}$. For the average relic $\nu$ momentum in the C${\nu}$B frame we have $|\vec{\tilde{p}}_{\nu_i,0}|\simeq 3.15\, T_{\nu,0}=0.5292 \times 10^{-3}\,\text{eV}$. 

In our computations we take the lightest neutrino mass to be $m_1=10^{-5}$ eV followed by $m_2=8.65\times10^{-3}$ eV and $m_3=5.013\times10^{-2}$ eV in the Normal Hierarchy scenario.

Going forward we make use of Eq. (\ref{Fluxavg}) and (\ref{Hel4}) to arrive at the following flux averages (in units of eV),
\begin{align}
&\Phi^1_i=\biggl\langle\frac{(S_{\nu_i}\cdot p_e)(S_e \cdot p_{\nu_i})}{E_{p_{\nu_i}}} \biggl\rangle =(-27048.6,-52.105,-9.023)\,,\\
& \Phi^2_i=\biggl\langle\frac{(p_e \cdot p_{\nu_i})(S_e \cdot S_{\nu_i})}{E_{p_{\nu_i}}} \biggl\rangle=(-27051.8,-8383.33,-48324.5)\,,
\end{align}
such that their difference is,
\begin{equation}\label{faB}
	\Delta\Phi^T_i=\biggl\langle\frac{(S_{\nu_i}\cdot p_e)(S_e \cdot p_{\nu_i})}{E_{p_{\nu_i}}}\biggl\rangle-\biggl\langle\frac{(p_e \cdot p_{\nu_i})(S_e \cdot S_{\nu_i})}{E_{p_{\nu_i}}} \biggl\rangle=(3.2, 8331.225, 48315.477)\,,
\end{equation}
which is the relevant quantity for our computation ($i$\,=\,1\,,\,2\,,\,3).
Now we have all the necessary ingredients to explicitly calculate the energy shift. Using the value of $G_F=1.1663787\times10^{-23} \text{eV}^{-2}$, we find that,
\begin{equation}
	\Delta E_e(0,s_e)= s_e \times 3.798 \times 10^{-37} \text{eV},
\end{equation}
and taking the difference of energies between $s_e=+1 ~\text{and} ~(-1)$ we get the final energy shift between the two electron spin states,
\begin{equation}
	\Delta E_e= 7.597 \times 10^{-37} \text{eV} \sim 10^{-36} \text{eV}.
\end{equation}
Hence, for our BP we see that the corresponding energy shift is of the order of $10^{-36} \text{eV}$. We further clarify that not all of our plots reach this energy scale.
As an aside, we also depict the value of the flux average difference after setting  $m_e=1$ (depicting the effect of the $e$ mass scale) which evaluates to,
	\begin{equation}\label{faeB}
		\Delta\Phi^T_i=(6.303\times10^{-6}, 0.016, 0.095)\,.
	\end{equation}
Comparing Eq. \eqref{faeB} with Eq. \eqref{faB} we see that the electron mass plays a crucial role in enhancing the flux average values. This phenomenon is naturally absent in the flux averages for SM and NSI sectors due to the absence of $p_e$. 
\subsection{$\Delta E_{\text{SM}}$}
We briefly depict the calculation of the energy shift for the pure Standard Model case with all BSM parameters set to zero. Again the general energy shift is given by,
\begin{align}\label{finalEmodB2}
	\Delta E^D_e (\vec{0},s_e)&=\frac{G_F}{\sqrt{2}}\sum_{i,s}\Biggr[\Bigl(n_{\nu}(\nu^D_{i,s})-n_\nu ({\Bar{\nu}}^D_{i,s})\Bigl)\Biggl\{\textcolor{red}{D^V_{ii}\biggl\langle\frac{S_e \cdot p_{\nu_i}}{E_{p_{\nu_i}}} \biggl\rangle}+\frac{2}{m_e}C^T_{ii}\biggl(\biggl\langle\frac{(S_{\nu_i}\cdot p_e)(S_e \cdot p_{\nu_i})}{E_{p_{\nu_i}}} \biggl\rangle\notag\\ &-\biggl\langle\frac{(p_e \cdot p_{\nu_i})(S_e \cdot S_{\nu_i})}{E_{p_{\nu_i}}} \biggl\rangle\biggl) \Biggl\}+\Bigl(n_{\nu}(\nu^D_{i,s})+n_\nu ({\Bar{\nu}}^D_{i,s})\Bigl)\Biggl\{\textcolor{red}{C^A_{ii}m_{\nu_i}
\biggl\langle\frac{S_e \cdot S_{\nu_i}}{E_{p_{\nu_i}}} \biggl\rangle} \Biggl\}\Biggr]\notag\\
&+\text{terms independent of}\:{S_e}\,,
\end{align}
where the terms depicted in red denote the SM and NSI contributions. For the purely SM case, 
\begin{align}
	D^V_{\alpha\beta}&=D^{V,SM}_{\alpha\beta}=-\delta_{\alpha e}\delta_{\beta e}+\frac{\delta_{\alpha\beta}}{2}=-C^A_{\alpha\beta}=-C^{A,SM}_{\alpha\beta}\notag\,,\\
	D^A_{\alpha\beta}&=D^{A,SM}_{\alpha\beta}=-\delta_{\alpha e}\delta_{\beta e}+\biggl(\frac{1}{2}-2\text{sin}^2 \theta_W\biggl)\delta_{\alpha\beta}=-C^V_{\alpha\beta}=-C^{V,SM}_{\alpha\beta}\notag\,,\\
&\text{and all other $C^T_{\alpha\beta}$ and $D^T_{\alpha\beta}=0$}\,,
\end{align}
which after going to the mass basis results in
\begin{equation}
	D^{V,SM}_{ii}=(-0.177,0.199,0.478)=-C^{A,SM}_{ii}\,.
\end{equation}
We take an $\mathcal{O}(10^{-2})$ asymmetry considering an excess of neutrinos over anti-neutrinos. In the C$\nu$B frame this asymmetry is $\Delta n_0 \simeq 3.156 \times 10^{-14}\, \text{eV}^3$. We then find the following Earth frame number densities (in the units of $10^{-13}\text{eV}^3$),
\begin{align}
	&n_\nu(\nu^D_{1,L})\simeq 4.643\,,~n_\nu(\nu^D_{2,L})\simeq 4.620\,,~ n_\nu(\nu^D_{3,L})\simeq 4.504\,, \notag\\
	&n_\nu(\nu^D_{1,R})\simeq 1.479\times 10^{-3}\,,~ n_\nu(\nu^D_{2,R})\simeq 2.421\times 10^{-2}\,,
 n_\nu(\nu^D_{3,R})\simeq 1.403\times 10^{-1}\,,\notag\\
	&n_\nu(\bar{\nu}^D_{1,R})\simeq 4.328\,,~n_\nu(\bar{\nu}^D_{2,R})\simeq 4.306\,,~n_\nu(\bar{\nu}^D_{3,R})\simeq 4.198\,,\notag\\
	&n_\nu(\bar{\nu}^D_{1,L})\simeq 1.378\times 10^{-3}\,,~n_\nu(\bar{\nu}^D_{2,L})\simeq 2.257\times 10^{-2}\,,~n_\nu(\bar{\nu}^D_{3,L})\simeq 1.307\times 10^{-1}\,.
\end{align}
The flux averages associated with the SM sector are given by,
\begin{align}
&\Phi^3_i=\biggl\langle\frac{S_e \cdot p_{\nu_i}}{E_{p_{\nu_i}}} \biggl\rangle =(-0.001,-0.0013,-0.0013)\,,\\ 
& \Phi^4_i=\biggl\langle\frac{S_e \cdot S_{\nu_i}}{E_{p_{\nu_i}}} \biggl\rangle=(-66.691,-1.893,-1.886)\,.
\end{align}
We can now explicitly calculate the energy shift for the purely SM scenario. Taking the difference of energies between $s_e=+1 ~\text{and} ~(-1)$ we get the final energy shift between the two electron spin states as,
\begin{equation}
	\Delta E_e= 2.331 \times 10^{-38} \text{eV} \sim 10^{-38} \text{eV}.
\end{equation}
Hence, for the Standard Model we see that the energy shift is of the order of $10^{-38} \text{eV}$ corresponding to an asymmetry of $\mathcal{O}(10^{-2})$. 

\bibliographystyle{JHEP}
\bibliography{newref.bib}

\end{document}